\documentclass[nohyper, 12pt,letterpaper]{JHEP3}
\usepackage[dvips]{epsfig}
\usepackage{amsfonts,amssymb}
\usepackage{multicol}

\newcommand{\Ub}{\bar{U}}
\newcommand{\Vb}{\bar{V}}
\newcommand{\Yb}{\bar{Y}}
\newcommand{\Zb}{\bar{Z}}

\newcommand{\hs}{\hspace{0.3 cm}}
\newcommand{\hsn}{\hspace{-0.135 cm}}

\newcommand{\mF}{\mathcal{F}}
\newcommand{\mW}{\mathcal{W}}\newcommand{\cW}{\mathcal{W}}

\newcommand{\onesusyadscft}{\mbox{$\cN=1$ AdS/CFT}}

\newcommand{\beq}{\begin{equation}}
\newcommand{\eeq}{\end{equation}}
\newcommand{\beqa}{\begin{eqnarray}}
\newcommand{\eeqa}{\end{eqnarray}}

\newcommand{\be}{\begin{equation}}
\newcommand{\ee}{\end{equation}}
\newcommand{\bea}{\begin{eqnarray}}
\newcommand{\eea}{\end{eqnarray}}
\newcommand{\ba}{\begin{eqnarray}}
\newcommand{\ea}{\end{eqnarray}}

\def\a{\alpha}
\def\b{\beta}

\newcommand{\cH}{\mathcal{H}}
\newcommand{\cN}{\mathcal{N}}
\newcommand{\cO}{{\cal O}}
\newcommand{\cP}{\mathcal{P}}
\newcommand{\cS}{{\cal S}}
\newcommand{\cK}{{\cal K}}
\newcommand{\cD}{{\mathcal D}}
\newcommand{\cDb}{{\bar{\mathcal D}}}

\newcommand{\ads}[1]{\mbox{${AdS}_{#1}$}}
\newcommand{\adss}[2]{\mbox{$AdS_{#1}\times {S}^{#2}$}}

\newcommand{\eg}{{\it e.g.}}

\newcommand{\commentout}[1]{}

\newcommand{\cL}{{\cal L}}

\newcommand{\cY}{\ensuremath{{\cal Y}}}

\newcommand{\Ypq}{\ensuremath{{\cal Y}^{p,q}}}

\newcommand{\half}{\ensuremath{\frac{1}{2}}}

\newcommand{\sql}{\ensuremath{\sqrt{\lambda}}}

\newcommand{\isql}{\ensuremath{\frac{1}{\sqrt{\lambda}}}}

\newcommand{\N}[1]{\ensuremath{\cN=#1}}

\newcommand{\ps}{\ensuremath{\partial_\sigma}}

\newcommand{\pt}{\ensuremath{\partial_t}}
\newcommand{\pta}{\ensuremath{\partial_{\tau}}}

\newcommand{\Pt}{\ensuremath{P_\theta}}
\newcommand{\Pa}{\ensuremath{P_\alpha}}
\newcommand{\Pps}{\ensuremath{P_\psi}}
\newcommand{\Pph}{\ensuremath{P_\phi}}
\newcommand{\Py}{\ensuremath{P_y}}

\newcommand{\tpsi}{\ensuremath{\tilde{\psi}}}

\newcommand{\ct}{\ensuremath{\cos\theta}}

\newcommand{\sqt}{\ensuremath{\sin^2 \theta}}

\newcommand{\bra}[1]{\mbox{$\langle #1 |$}}
\newcommand{\ket}[1]{\mbox{$| #1 \rangle$}}

\newcommand{\tr}{\mbox{tr}}

\def\lbldef#1#2{\expandafter\gdef\csname #1\endcsname {#2}}

\newcommand{\ds}{d \sigma}

\title{Semiclassical \hsn strings \hsn in \hsn Sasaki-\hspace{-0.08cm}Einstein \hsn manifolds \\
and long operators in \N{1} gauge theories
\\
}

\author{
Sergio Benvenuti \thanks{E-mail: sergio.benvenuti@sns.it} \\
   Scuola Normale Superiore, Pisa, \\
   and INFN, Sezione di Pisa, Italy.
}

\author{
Martin Kruczenski  \thanks{E-mail: martink@brandeis.edu} \\
  Department of Physics, Brandeis University \\
  Waltham, MA 02454, USA.
}

\abstract{
 We study the AdS/CFT relation between an infinite class of
5-d \cY$^{p, q}$ Sasaki-Einstein metrics and the corresponding quiver theories.
The long BPS operators of the field theories are matched to massless
geodesics in the geometries, providing a test of
AdS/CFT for these cases. Certain small fluctuations (in the BMN sense)
can also be successfully compared.

 We then go further and find, using an appropriate limit, a reduced action, first order in
time derivatives, which describes strings with large R-charge. In the field theory we consider holomorphic
operators with large winding numbers around the quiver and find, interestingly, that, after certain simplifying assumptions,
they can be described effectively as strings moving in a particular metric. Although not equal,
the metric is similar to the one in the bulk.
We find it encouraging that a string picture emerges directly from the field theory
and discuss possible ways to improve the agreement.
}


\preprint{\tt{BRX TH-564} \\
          \tt{hep-th/0505046}  }

\begin{document}

\section{Introduction} \label{intro}

 The AdS/CFT correspondence \cite{malda} gave a precise example of the conjectured relation \cite{largeN} between
the large $N$ limit of gauge theories and string theory. The most studied model is \N{4} SYM, with gauge group
$SU(N)$ and coupling $g_{\rm YM}$. In its simplest form, the correspondence establishes that, in the large $N$-limit,
keeping $\lambda= g_{\rm YM}^2 N$ fixed, this theory is the same as free IIB strings on \adss{5}{5}, with $N$ units
of RR 5-form flux. The radius $R$ of \adss{5}{5} is given by $R/l_s= \lambda^{\frac{1}{4}}$ where $l_s$ is the
string length. The effective string tension is therefore $\lambda^{-\half}$. In the limit $\lambda\rightarrow\infty$
the worldsheet theory becomes classical and can be easily studied. On the other hand, the field theory simplifies
in the opposite limit, $\lambda\rightarrow 0$, since then it becomes perturbative. This makes the correspondence
very powerful but at the same time difficult to study. In particular, it does not elucidate the point of how a
string description can emerge from a field theory (see \cite{POL} for a discussion). An important step in that
direction was made in \cite{bmn} where it was shown  how this correspondence can be established for certain
ultrarelativistic strings, i.e.  strings whose  kinetic energy is much larger than their mass. In a related paper \cite{gkp},
the string side of the calculation was understood from a semi-classical point of view which allowed a generalization
of \cite{bmn} to multispin string states in \cite{ft1,ft2} (see
\cite{tse1,tse2} for a review and \cite{dev} for previous related work).
The calculation can also be generalized thanks to the observation of \cite{mz1} that the one-loop scalar
dilatation operator can be interpreted as a Hamiltonian of an integrable $SO(6)$ spin chain. Using a Bethe ansatz
method to solve a subsector of the spin chain, in \cite{bmsz,bfst} a remarkable agreement was found between energies
of  various string solutions and eigenvalues  of the dilatation operator representing dimensions of particular SYM
operators. Moreover, integrable structures appear for certain rigid-shape rotating string configurations
\cite{afrt,art} and can be mapped \cite{as,Min2} to the integrable structure of the spin chain.

Another step was made  in  \cite{kru}, where it was shown that one can take the ultrarelativistic limit directly
in the string action\footnote{See \cite{mik,Gor} for alternative
approaches and \cite{mateos} for a discussion of supersymmetry in
  the ultrarelativistic limit.}.  The resulting, reduced action is a sigma model which
turns out to be precisely the semiclassical coherent state action describing the field theory
spin chain (in an $SU(2)$ subsector). This makes obvious how a sigma model description of
operators can emerge from a field theory as an effective description of very long operators.
These ideas can be cast also in the framework of integrable models as later shown in
\cite{kmmz}. The results are also useful in understanding higher orders of the semi-classical approximation \cite{krt},
other subsectors \cite{lopez,ST,DR,KT,Hernandez:2004kr,Kruczenski:2004wg,Stefanski:2005tr,Bellucci:2005vq} including open strings \cite{open}, and also
quantum corrections \cite{1loop}. It could be useful also in understanding $1/N$ corrections as the ones discussed 
in \cite{Peeters:2005pb}.

Moreover, the relation between spin chains and gauge theory is
quite generic and in fact it was already noted in QCD \cite{QCD}
implying that these ideas have wide applicability. Therefore, it
is natural to wonder if they can be extended to other examples of
the AdS/CFT correspondence, for example, with less than four
supersymmetries. Examples with at least one supersymmetry are
generically\footnote{This does not include the Maldacena-Nu\~nez
solution \cite{MN} which uses a different approach. See
\cite{Butti:2004pk,
  Gubser:2004tf} for more generic solutions.} given by IIB backgrounds of the form \ads{5}$\times X^5$,
where $X^5$ is a five dimensional Sasaki-Einstein manifold. The dual superconformal theories are quivers, arising
from the low energy excitations of D3-branes at Calabi--Yau singularities. Until recently, the only examples where
the metric on $X_5$ was explicitly known were the homogeneous manifolds $S^5$ and $T^{1,1}$. The latter case, discussed
in detail by Klebanov and Witten \cite{kw}, gives the paradigmatic example of \onesusyadscft.

 However, one year ago, Gauntlett, Martelli, Sparks and Waldram found an infinite class of inhomogeneous
Sasaki--Einstein metrics on $S^2 \times S^3$ \cite{Gauntlett:2004zh, Gauntlett:2004yd, Gauntlett:2004hh}
which are labeled by two integers $0 \leq q \leq p$, and are usually called \cY$^{p,q}$ metrics. The corresponding Calabi--Yau cones,
are \emph{toric}, meaning that there is an effectively acting $U(1)^3$
isometry. The toric description of the geometries was given in \cite{Martelli:2004wu} which allowed \cite{Benvenuti:2004dy} to find the superconformal gauge theories
dual to Type IIB on \mbox{$AdS_5 \times \cY^{p,q}$}.

Once the superconformal field theories are known, it's possible to
 compute the anomalous dimensions of the
 chiral fields applying the general  $a$-maximization technique of
 \cite{intriligator03}, which relies on general properties of supersymmetric theories \cite{Anselmi:1997am} and works independently of AdS/CFT.
 These anomalous dimensions are directly related, in the
supergravity dual, to the volume of the dual Sasaki--Einstein
manifold, as well as the volumes of supersymmetric submanifolds.
In fact, \cite{Martelli:2005tp} later found the geometric analog
of $a$-maximization, i.e. a general way of computing these volumes
for toric Sasaki--Einstein manifolds in any dimension, bypassing
the need of an explicit knowledge of the metric.

 In a further development \cite{Hanany:2005ve}, a relation was pointed out between toric quivers and
dimers, that leads \cite{Franco:2005rj} to a general method for obtaining the corresponding brane setups
\cite{Hanany:1996ie}. This 'brane tiling' technique is connected to a correspondence between the statistical mechanics
of dimers and topological strings on Calabi-Yau's \cite{Okounkov:2003sp}, and significantly generalizes previously
known similar constructions \cite{pqwebs, braneboxes}.

 In the present paper similar periodic representations of quivers are considered, in particular, the
full mesonic chiral ring of toric gauge theories is naturally encoded in one-cycles of the torus where
the quiver itself is drawn.

Another generic feature of quivers associated to toric geometries is that they always admit an
exactly marginal deformation analogous to the  $\beta$-deformation of
\N{4} SYM \cite{Leigh:1995ep}, as was shown in \cite{Benvenuti:2005wi}, using the techniques of \cite{Leigh:1995ep,
  Kol:2002zt}. This deformation leaves the toric $U(1)^2_F \times
U(1)_R$ isometry untouched. In \cite{Lunin:2005jy} a very
interesting way of constructing the gravity side of this kind of
deformations has been found. The semiclassical sector of the
correspondence \cite{Frolov:2005ty} and integrability properties
\cite{ Frolov:2005ty2} have been studied.

The knowledge of a general class of geometries should allow the
construction of non-conformal examples of the correspondence, in the
spirit of the Klebanov-Strassler solution \cite{KlebanovProgram}. Progress in this direction
was done in \cite{Herzog:2004tr, Pal:2005mr, Burrington:2005zd,  Berenstein:2005xa, franco2005}.

 In this paper we are interested in improving the understanding of the correspondence in these new examples.
 We proceed in steps. First we establish a correspondence between massless geodesics and large R-charge
chiral primary operators in the field theory. After that, following \cite{bmn}, we consider excited strings
whose mass is small compared to their kinetic energy. When the string has few excitations (of certain types)
we can find the corresponding operators in the field theory. For a large number of excitations, however, we
need a way to obtain a spin chain description of the operators. This is difficult since the theory is not in
a perturbative regime. We content ourselves with analyzing the mixing between operators induced by the
superpotential and show that they lead to a sigma model action which has similar properties as an action that
can be derived directly form the string side of the correspondence. The sigma models are not the same but we
suggest that they should have the same infrared limit (in the worldsheet sense). So, we can argue that we indeed
were able to find a string action emerging from the field theory. The mapping from one side of the correspondence
to the other is that long paths in the quiver correspond to strings. If one draws the paths in a torus then
the direction in which the path moves is directly related to the position of the string in the bulk.

 The organization of this paper is as follows: In sec. \ref{strings} we analyze strings moving in the \cY$^{p, q}$
manifolds. We find massless geodesics paying special attention to the ones corresponding to chiral primaries
(long) operators in the quivers.  We then consider semiclassical fast moving strings, in the limit of \cite{kru}.
The Sasaki-Einstein geometry seen by these strings is naturally  parameterized by a non relativistic effective action,
that keeps all the information about the Sasaki--Einstein metric.

 In sec. \ref{Field theory} the BPS sector of the field theory is analyzed. The full mesonic chiral ring is constructed
for a general \cY$^{p, q}$ gauge theory, exploiting general features of toric quivers. We reobtain the results on BPS
geodesics and find the natural ranges of the  coordinates parameterizing the \cY$^{p, q}$ manifolds.

  In order to reconstruct the full string background, it is necessary
 to go beyond the BPS sector and consider chiral non BPS operators. In sec. \ref{lambda0} we show the
 existence of a special point on the
 conformal manifold where some coefficents of the superpotential vanish.
  At that point the chiral ring is enhanced and a large class
 of holomorphic operators becomes BPS\footnote{This is the class of operators that we call "holomorphic sector".
  In the case of \N{4} this sector is the well known $SU(3)$ subsector. We don't discuss
the closure of the sector in the gauge theories, but we expect
that for long operators this sector becomes closed, as in \N{4}
\cite{Minahan:2004ds}.}. For long operators, non
 trivially, this class includes the extended semiclassical strings, as
 can be expected from the string side. The existence of a special
 point with enhanced chiral ring and symmetry (noticed also in \cite{Benvenuti:2005wi} in some particular examples) is
 an exact result that we expect to be a general feature of
 superconformal quiver gauge theories and thus of AdS/CFT with critical string theory.

In sec. \ref{spin chain action} we construct a spin chain
Hamiltonian for the \cY$^{p, q}$ quivers, considering a simplified
approach consisting in studying the mixings between chiral
operators induced by the superpotential terms in the Lagrangian.
In this way, we are able to reconstruct an $S^2 \times S^3$
geometry from the chiral semiclassical operators in the field
theory. Even with the mentioned simplifications, the metric found
is very similar to the original Sasaki--Einstein metric. More
precisely, we find a K\"ahler metric on the base but the metric is
not Einstein. We suggest that the metric may flow to an Einstein
metric in the infrared of the world-sheet.

In sec. \ref{generic}, which can be read independently of sec.
\ref{lambda0} and \ref{spin chain action}, we extend the results
of sec. \ref{Field theory} in a different direction. Instead of
considering extended strings we consider non-BPS massless
geodesics. We find a class of operators that we conjecture to
correspond to a generic non BPS geodesic, and test the idea for
massless strings moving along a small perturbation of a BPS
geodesic. For short operators, this leads to a proposal for the,
generically non protected, operators dual to all supergravity
states, i.e. generic Kaluza-Klein harmonics on the transverse
Einstein manifold.

Finally we give our conclusions in section \ref{conclu}.


\section{Strings moving in the $\Ypq$ manifold} \label{strings}

 In this section we study semiclassical strings moving in the $\ads{5}\times\Ypq$ manifold whose metric is \cite{Gauntlett:2004yd}:
\beqa
 ds^2&=& - dt^2\, \cosh^2\!\rho + d\rho^2 + \sinh^2\!\rho\, d\Omega_3^2 + ds_{p,q}^2 \label{metric1}\\
 ds_{p,q}^2 &=& \frac{1-y}{6} \left(d\theta^2+\sqt d\phi^2\right)+ \frac{dy^2}{6 p(y)} +
                \frac{q(y)}{9} \left(d\psi - \ct d\phi\right)^2  \label{metric2} \\
            & & + w(y)\left[d\alpha+f(y)\left(d\psi-\ct d\phi\right)\right]^2
\label{metric}
\eeqa
with the functions
\beq
w(y)= 2\,\frac{a - y^2}{1 - y},\;\;\; q(y)=\frac{a - 3 y^2 + 2 y^3}{a - y^2},
\;\;\; f(y)=\frac{a-2y+y^2}{6(a-y^2)} ,
\label{wqfdef}
\eeq
and
\beq
p(y) = \frac{w(y) q(y)}{6} = \frac{a-3y^2+2y^3}{3 ( 1 - y )}
\label{pdef}
\eeq
The coordinates span the range:
\beq
0\le \theta \le \pi, \ \ \ 0\le \phi \le 2\pi , \ \ \ 0\le \psi \le 2\pi
 , \ \ \ 0\le \alpha \le 2\pi \ell , \ \ \ y_1\le y \le y_2
\label{arange}
\eeq
The constant $a$ appearing in the metric as well as the constants $y_{1,2}$ and $\ell$ which determine
the range of variation of the coordinates can all be written in terms of the integers
$p$ and $q$ that define the manifold:
\beqa
 y_{1,2} &=& \frac{1}{4p}\left(2p \mp 3q -\sqrt{4p^2-3q^2}\right) \\
 \ell &=& \frac{q}{3q^2-2p^2+p\sqrt{4p^2-3q^2}} \label{defell}\\
 a &=& 3y_1^2-2y_1^3
\label{yla}
\eeqa
 An important point is that $y_{1,2}$ are zeros of the function $p(y)$ appearing in the metric.
We note for further use that there is a third zero of $p(y)$ given by $y_3=\frac{3}{2}-y_1-y_2$.
Various useful properties of these functions and the metric can be found in the original
paper \cite{Gauntlett:2004yd} and are collected in an Appendix for completeness.

\subsection{Massless geodesics} \label{massless}
We consider massless geodesic in the reduced metric
\beq\label{redgeod}
ds^2 = -dt^2 + ds_{p,q}^2 = -dt^2 + g_{ab} dx^a dx^b
\eeq
where $ds_{p,q}^2=g_{ab} dx^a dx^b$ is the metric of the Sasaki-Einstein manifold and $t$ is the global
time in \ads{5}. The massless point-like string is sitting at $\rho=0$ in the metric (\ref{metric1}) and
the motion is only in the internal manifold.

The action for the motion of a point-like string is
\beq
S = \frac{\sqrt{\lambda}}{2} \int d\tau \left( -\dot{t}^2 + g_{ij} \dot{x}^a \dot{x}^b \right)
\eeq
where $\sql=(R/l_s)^2$ is the effective string tension. We include it for completeness but the
results do not depend on the tension since the strings are point-like.
We need to solve the equations of motion subject to the constraint
\beq
-\dot{t}^2 + g_{ab} \dot{x}^a \dot{x}^b = 0
\label{constraint}
\eeq
The equation of motion for $t$ is solved by $t=\kappa \tau$ and therefore the action reduces
\beqa
S &=& \int d\tau \cL = \frac{\sql}{2} \int d\tau \left( g_{ab} \dot{x}^a \dot{x}^b \right) \\
 &=& \frac{\sql}{2} \int d\tau \left\{ \frac{1-y}{6} \left(\dot{\theta}^2 + \sin^2\theta\dot{\phi}^2\right) +
\frac{1}{w(y)q(y)} \dot{y}^2 + \frac{q(y)}{9} \left(\dot{\psi}-\cos\theta\dot{\phi}\right)^2 \right. \\
&& \left. + w(y) \left[\dot{\alpha} +f(y)\left(\dot{\psi}-\cos\theta\dot{\phi}\right) \right]^2 \right\}
\eeqa
namely free motion in the Sasaki-Einstein manifold. The momentum $P_t$ conjugate to $t$ is the energy of the string
and therefore is equal to the conformal dimension $\Delta$ of the dual operator:
\beq
\Delta = P_t = \sql \kappa
\label{Dkappa}
\eeq
We can also introduce the other conjugate momenta as
\beq
p_a = \frac{\partial \cL}{\partial \dot{x}^a}
\eeq
and the Hamiltonian which is given by
\beq
 H = \frac{1}{2} g^{ab} p_a p_b
\eeq
{}From the action we see immediately that the momenta $P_{\phi}$, $P_{\psi}$ and $\Pa$
are conserved quantities. This is a consequence of the $SU(2)\times U(1) \times U(1)$ isometry
since $P_{\phi}$ is the third component of the $SU(2)$ angular momentum and
$P_{\psi}$, $\Pa$ are associated to the $U(1)$ factors. There is a further conserved quantity
corresponding to the total $SU(2)$ angular momentum given by:
\beq
J^2 = \Pt^2 + \frac{1}{\sqt} \left(\Pph + \ct \Pps\right)^2 + \Pps^2
\label{J}
\eeq
The momenta in terms of the velocities are given by
\beqa
\isql\,P_y &=& \frac{1}{6 p(y)} \dot{y} \\
\isql\,P_{\theta} &=& \frac{1-y}{6} \dot{\theta} \\
\isql\,\left( P_{\phi} + \cos\theta P_{\psi}\right)&=& \frac{1-y}{6} \sin^2\theta \dot{\phi}  \\
\isql\,\left(P_{\psi} - f(y) \Pa \right)  &=& \frac{q(y)}{9} \left(\dot{\psi} - \cos\theta \dot{\phi}\right) \\
\isql\,\Pa &=& w(y) \left(\dot{\alpha} + f(y)  \left(\dot{\psi} - \cos\theta \dot{\phi}\right)\right)
\eeqa
In terms of the momenta the Hamiltonian can be written as
\beqa
\sql\, H &=& \frac{\lambda}{2} \kappa^2 = \half \Delta^2
    = \frac{1}{2} \left\{ 6 p(y) P_{y}^2 + \frac{6}{1-y} \left(J^2-P_{\psi}^2\right)
       + \frac{1-y}{2(a-y^2)} \Pa^2\right. \\
  && \left.  + \frac{9(a-y^2)}{a-3y^2+2y^3}\left(P_{\psi}- \frac{a-2y+y^2}{6(a-y^2)}\Pa\right)^2\right\}
\eeqa
 where we also used the constraint (\ref{constraint}) to relate $H$ to $\kappa$ and further used (\ref{Dkappa}) to
relate $\kappa$ and $\Delta$. As expected the relation between the conformal dimension $\Delta$ and the momenta
does not involve the tension $\sql$.

 The only non trivial equation of motion we need to solve now is that of $y(\tau)$ which is simply a one dimensional
motion in a potential as follows from the conservation of $H$ and the fact that $P_{y} \propto \dot{y} / p(y)$. Before
proceeding it is useful to introduce the R-charge:
\beq
Q_R = 2P_{\psi}-\frac{1}{3} \Pa
\label{QRdef}
\eeq
which gives,  after some algebra,
\beqa \label{fundgeod}
 \Delta^2 &=&  \left(\frac{3}{2}\, Q_R\right)^2  +
  \frac{1}{6 p(y)} \left(\Pa+3\,y\,Q_R\right)^2 \\
       &&   + 6 p(y) P_{y}^2 + \frac{6}{1-y} \left(J^2-P_{\psi}^2\right) \nonumber
\eeqa
where we used the function $p(y)$ that was defined in (\ref{pdef}).
As we said this last equation should be understood as an equation of motion for $y(\tau)$.

The full set of geodesics moving only in the transverse SE manifold is completely described by eq. (\ref{fundgeod}). We note that the set of geodesics on a five dimensional manifold is itself a manifold with eight dimensions; in the case of $S^5$, for instance, this set is the manifold $SO(6) / ( SO(2) \times SO(4) )$. Since from (\ref{J}) $J^2\ge \Pps^2$, all solutions have $\Delta \ge \frac{3}{2} Q_R$. We want now to restrict to solutions where this bound is saturated. These geodesics correspond to chiral primary, or BPS, operators that will be analyzed in the next section. From (\ref{fundgeod}) it is clear than in order to have $\Delta = \frac{3}{2} Q_R$ we must require
\beq
\Py=0, \ \ \ J^2=\Pps^2,  \label{cond1}
\eeq
 The first equation implies $y=y_0$ is constant. The constant $y_0$ should be set to the minimum
of $\Delta^2$, namely
\beq
y_0 = -\frac{\Pa}{3Q_R} \label{cond2}
\eeq
 This ensures that the equation of motion for $y$ is satisfied and at the same time implies $\Delta=\frac{3}{2}Q_R$.
The restriction however is that, to obtain a geodesic, we need:
\beq
 y_1 \le -\frac{\Pa}{3Q_R} \le y_2
\eeq
 In this sense, $y_{1,2}$ can be thought as defining the range of variation of $\Pa/Q_R$.

 To summarize, for all these BPS geodesics we obtain:
\beq
 \Pa = -3y_0 Q_R, \ \ \ J = \half (1-y_0)Q_R
\label{PaJ}
\eeq
and the $y_0$ independent relations
\beq
\Delta = \frac{3}{2} Q_R , \ \ \ Q_R = 2 J -\frac{1}{3} \Pa.
\label{DQ}
\eeq
The last equality follows from the definition of $Q_R$, namely eq.(\ref{QRdef}), and the
fact that $J=P_{\psi}$ for these geodesics. Together with the first relation in (\ref{PaJ})
it implies the second one, namely $ J = \half (1-y_0)Q_R$.

Using the definitions of the momenta in terms of the velocities one can see that the geodesics in question are simply given by
\beq
y=y_0, \ \  \theta=\theta_0 , \ \ \phi =\phi_0, \ \ \dot{\alpha} + \frac{1}{6} \dot{\psi} =0
\label{geod1}
\eeq
which suggests introducing a new angle $\beta$ through
\beq
\beta = 6\, \alpha + \psi , \ \ \ \tilde{\psi}=\psi
\eeq
 This implies
\beqa
P_{\beta} &=& \frac{1}{6}\, \Pa \\
P_{\tilde{\psi}} &=& \Pps - \frac{1}{6} \Pa = \half Q_R
\eeqa
 Now, the geodesics are such that $\dot{\beta}=0$. Note also that $\tilde{\psi}=\psi$ is now conjugate
to the R-charge.

Concluding, a four dimensional subset of geodesic is BPS, and corresponds, as expected, to point-like strings moving
only along the R-charge direction, or $\tilde{\psi}$ direction. In section \ref{Field theory} we reconstruct these
BPS geodesics from chiral primaries in the quivers. In section \ref{generic} we will study small deviations from
the BPS case, corresponding to \mbox{$\Delta > \frac{3}{2}\, Q_R$}.

\subsection{Reduced action for strings with large R-charge} \label{reducedaction}
In this section we consider classical strings which move with large angular momentum
corresponding to field theory operators with large R-charge. Such
strings move fast in the $\psi$ direction as in the previous section
but now we do not use the approximation that the string is small. Any five-dimensional Sasaki--Einstein metric can be written in the following form
\beq
ds^2 = -dt^2 + \frac{1}{6} g_{ij} dx^i dx^j + \frac{1}{9} (d\psi + A_i dx^i)^2
\eeq
where $g_{ij}$ is a local K\"ahler--Einstein metric on the base, parameterized by the coordinates $x_i$.
Both $g_{ij}$ and $A_i$ depend only on the four coordinates $x_i$. The external derivative of the one
form $A_i dx^i$ is proportional to the K\"ahler form of the four dimensional base. It also completely
specifies the K\"ahler form of the Calabi-Yau cone over the SE manifold.

Now we introduce a coordinate $\psi_1 = \psi - 3 t$. The metric becomes
\beq
ds^2 = \frac{2}{3} dt(d{\psi_1}+A_i dx^i) +\frac{1}{9}(d{\psi_1}+A_idx^i)^2 + \frac{1}{6} g_{ij} dx^i dx^j
\eeq
If we choose $t=\kappa \tau$ we can write the Polyakov action
\beqa
 S &=& \half \int \frac{2}{3} \kappa(\pta{\psi_1}+ A_i \pta{x}^i) + \frac{1}{9} (\pta{\psi_1}
       +A_i\pta{x}^i)^2 + \frac{1}{6} g_{ij} \pta{x}^i \pta{x}^j \\
   &&  - \frac{1}{9} (\ps{\psi_1}+A_i \ps {x}^i)^2 - \frac{1}{6} g_{ij} \ps{x}^i \ps{x}^j
\eeqa
and the conformal constraints
\beqa
0&=&  \frac{1}{3} \kappa (\pta{\psi}_1 + A_i \pta{x}^i)
          + \frac{1}{9} (\pta{\psi_1}+A_i\pta{x}^i)^2 + \frac{1}{6} g_{ij} \pta{x}^i \pta{x}^j \\
 && + \frac{1}{9} (\ps{\psi_1}+A_i\ps{x}^i)^2 + \frac{1}{6} g_{ij} \ps{x}^i \ps{x}^j \\
0 &=&  \frac{2}{3} \kappa(\ps{\psi_1}+A_i \ps x^i) +\frac{1}{9}(\pta{\psi}_1+A_i\pta{x}^i)(\ps{\psi_1}+A_i\ps x^i) + \frac{1}{6} g_{ij} \pta{x}^i \ps x^j
\eeqa
In this system of coordinates the string moves slowly (which means it moves almost at the speed of light in the original
ones). We therefore consider the limit~\cite{kru}
\beq
 \pta{X} \rightarrow 0, \ \ \ \kappa\rightarrow\infty, \ \ \ \kappa \,
 \pta{X} \ \mbox{fixed}
\eeq
where $X$ denotes all coordinates, $\psi_1$ and $x^i$.

In that limit the second conformal constraint reduces to
\beq
\ps{\psi_1}+A_i \ps x^i = 0 \label{gengaugech}
\eeq
Taking the limit in the action and using the constraint we get
\beq\label{genredaction}
S =  \int \frac{1}{3} \kappa (\pta\psi_1+ A_i \pta{x}^i)  - \frac{1}{12} g_{ij} \ps{x}^i \ps{x}^j
\eeq
which is the final form of the reduced action describing strings with
large R-charge.

We can now specify this general derivation to our case of interest, the \cY$^{p, q}$ metrics. Using the coordinates discussed at the end of the previous subsection, $(\theta, \phi, y, \beta)$, the local K\"ahler--Einstein metric $g_{ij}$ and the $U(1)$-fibration $A_i$ are
\beqa\label{localmetric}
 g_{ij} dx^i dx^j &=&  (1-y) (d\theta^2 +\sin^2\theta d\phi^2) + \frac{dy^2}{p(y)}
  + p(y) (d\beta-\cos\theta d\phi)^2 \\
  A_i dx^i &=& - y d\beta-(1-y)\cos\theta d\phi\label{localoneform}
\eeqa
It should be noticed that the metric (\ref{localmetric}) is valid only
locally, and has orbifold singularities at the zeros of
$p(y)$.\footnote{We note also that the one form $A_i dx^i$ does not
  depend on $p(y)$; as a consequence the K\"ahler
forms of the four dimensional base and of the metric cone over the five-manifolds do not depend on the precise form of $p(y)$. In sec. \ref{spin chain action} we derive a metric from the spin chain which differs from (\ref{localmetric}) only in the form of $p(y)$ and for $A_i$ gives the same result as (\ref{localoneform}).} The constraint (\ref{gengaugech}) becomes
\beq
 \ps\psi_1 - \cos\theta\ps\phi = y\left(\ps\beta-\cos\theta\ps\phi\right)
\label{gaugech}
\eeq
and the effective action (\ref{genredaction}) takes the explicit form
\beqa
S &=& \sqrt{\lambda}\int \frac{1}{3} \kappa (\pta\psi_1 - y \pta\beta-(1-y)\cos\theta \pta\phi) \\
   &&  - \frac{1}{12} \left[(1-y) \left((\ps\theta)^2 +\sin^2\theta (\ps\phi)^2\right) + \frac{(\ps y)^2}{p(y)}   + p(y) (\ps\beta-\cos\theta \ps\phi)^2 \right]
\label{redaction}
\eeqa
where we restored the factor $(R/l_s)^2 = \sqrt{\lambda}$ in front of the action.
We can immediately identify the following conserved quantities
\beqa
 \cP_{\psi_1} &=&   \frac{1}{3} \sqrt{\lambda}\kappa\ \int\ds  = \frac{2\pi}{3} \sqrt{\lambda}\kappa \\
 \cP_{\beta} &=& - \frac{1}{3} \sqrt{\lambda}\kappa \int \ds\ y \\
 \cP_{\phi} &=&  - \frac{1}{3} \sqrt{\lambda}\kappa \int \ds\ (1-y) \cos\theta \label{Pphi}\\
 \cH &=& \frac{\sqrt{\lambda}}{12} \int\ds\ \left[(1-y)\left((\ps\theta)^2
                                 +\sin^2\theta (\ps\phi)^2\right) + \right. \\
    && \left. \phantom{aaaaaaaaa} \frac{(\ps y)^2}{p(y)} + p(y) (\ps\beta-\cos\theta \ps\phi)^2 \right]
\eeqa
where $\cP_{\psi_1}$ is (half) the R-charge,  $\cP_{\beta}$, $\cP_{\phi}$, are the $U(1)_F$ and the third component
of the $SU(2)$ charges respectively. Finally $\cH$ is the Hamiltonian which corresponds to
$\Delta-\frac{3}{2}Q_R$  in the field theory. Furthermore, if we remember that $P_{\psi_1}=P_{\tpsi}$, we see that the relations (\ref{PaJ}),(\ref{DQ}) are satisfied at each value of $\sigma$ implying that each point of the string moves approximately along a BPS geodesic. As expected, $\cH$ vanishes precisely when all the four local coordinates do not depend on $\sigma$. In this case one recovers the results of the previous subsection. It is useful to note that if we use the coordinate
$t=\tau/\kappa$ and replace $\kappa$ by the R-charge $Q_R =\half \cP_{\psi_1}$ the reduced action can be written as
\beqa
S &=& \frac{Q_R}{4\pi} \left\{ \int (\pt\psi_1 - y \pt\beta-(1-y)\cos\theta \pt\phi)\right. \\
   && \left. - \frac{4\pi^2}{9} \frac{\lambda}{Q_R^2} \left[(1-y) \left((\ps\theta)^2 +\sin^2\theta (\ps\phi)^2\right) + \frac{(\ps y)^2}{p(y)}   + p(y) (\ps\beta-\cos\theta \ps\phi)^2 \right] \right\} \nonumber
\label{redactionf}
\eeqa
 We see that the corrections introduced by a sigma dependence of the coordinates, namely for an extended string, are small
for large R-charge as expected. The result is valid for large $\lambda$ but a naive extrapolation to $\lambda=0$ suggests
that the corrections vanish in that point, a fact that we use later.

 We conclude this subsection by noting that one can write the reduced action (\ref{redaction}) in the following general form:
\beq
 S = -i\kappa\int \left(\dot{z}^a\partial_a K -\dot{\bar{z}}^{\bar{a}}\partial_{\bar{a}}K\right)
 - \half \int \partial_{a\bar{b}}K\, \ps z^{a}\ps \bar{z}^{\bar{b}}
\label{Kaction}
\eeq
where we introduced two complex variables $z^{a=1,2}$ and a K\"ahler
potential $K(z^1\bar{z}^{\bar{1}}+z^2\bar{z}^{\bar{2}})$. In terms of the original variables they
are:
\beqa
 z_1 &=& \sin(\frac{\theta}{2})\, e^{-i\half(\beta-\phi)}\, \prod_{i=1}^{3} |y-y_i|^{\frac{1}{4y_i}} \\
 z_2 &=& \cos(\frac{\theta}{2})\, e^{-i\half(\beta+\phi)}\, \prod_{i=1}^{3} |y-y_i|^{\frac{1}{4y_i}} \\
\eeqa
 where $y_1<y_2<y_3$ are the three roots of $p(y)=0$ already introduced in (\ref{yla}). This relation defines
complex coordinates only locally since for example the periodicity of $\beta$ is not $2\pi$.
 We see that $\rho=z^1\bar{z}^{\bar{1}}+z^2\bar{z}^{\bar{2}}$ is a function of $y$ only. This means that
the (local) K\"ahler potential is also a function of $y$ and turns out to be given by
\beq
K = \frac{1}{6}\sum_{i=1}^3 \frac{1-y_i}{y_i}\ln|y-y_i|
\eeq
 With these definitions it easy check that (\ref{Kaction}) is equivalent to (\ref{redaction}). In doing so
it is useful to note that
\beqa
\frac{\partial K}{\partial y} &=& -\frac{(1-y)}{3 p(y)} \\
\frac{\partial \rho}{\partial y} &=& - \frac{\rho}{p(y)}
\eeqa
 The form of the action (\ref{Kaction}) means simply that the base of the Sasaki-Einstein
manifold is locally K\"ahler with complex coordinates $z_{1,2}$ and K\"ahler potential $K$. The fact
that $K$ depends only on $\rho=z^1\bar{z}^{\bar{1}}+z^2\bar{z}^{\bar{2}}$ means that there is an
$U(2)=SU(2)\times U(1)$ isometry. Actually this fact supplemented by the condition that the metric is
Einstein completely determines the reduced action (up to the constants $y_{1,2}$ and the couplings).

\section{The correspondence in the BPS sector} \label{Field theory}
We now want to study the various operators dual to the semiclassical strings moving on the \cY$^{p, q}$ manifolds. Since we consider strings without AdS angular momenta, the operators are scalars constructed only with the matter bifundamental fields. Moreover, if the string is moving fast only along the $\psi$ direction, as is the case for strings described by the effective action of Sec. \ref{reducedaction}, the operators will be holomorphic, or chiral, i.e. products of chiral bifundamentals.

In this section we restrict to chiral primaries, or BPS operators. We will at first focus on the generators of the mesonic chiral ring. 'Mesonic' means that these operators are constructed taking the trace of products of bifundamental fields; in order to be gauge invariant each of these operators has to correspond to a loop in the quiver. Then we describe a generic BPS operator. This study reobtains our previous geometric results on BPS geodesics, and constitutes a first step towards the description of the holomorphic operators dual to semiclassical extended strings. Even if we consider the \cY$^{p,q}$ models, we will uncover generic features of toric quivers.

We refer to \cite{Benvenuti:2004dy} for the description of how the superconformal field theories are constructed. Table \ref{charges} gives the values of the R-charges $Q_R$, the $SU(2)$-spin $J$ and the $U(1)$ flavor charges $Q_F$ for all the bifundamental fields present in a generic \cY$^{p,q}$ quiver\footnote{In the case of \cY$^{2,1}$, for which the superconformal field theory was constructed in \cite{Feng:2000mi}, the values of the R-charges were computed in \cite{Bertolini:2004xf}.}. As is well known, for chiral operators there is a simple relation between the R--charge $Q_R$
and the scaling dimension $\Delta$: $\Delta = 3/2 Q_R$. Since the R-charges
add under multiplication of chiral operators, the knowledge of the R-charge of the four types of bifundamental fields, $Y, Z, U^\a$ and $V^\a$, suffices to determine the R-charge of all holomorphic operators and the scaling dimension of all chiral primaries.

\begin{table}[!h]\begin{center}
$$\begin{array}{|c|c|c|c|c|}  \hline
\hs\mathrm{Field}\hs&\hs\mathrm{number}\hs&   Q_R  & \hs U(1)_B \hs & \hs Q_F \hs\\ \hline\hline
Y             & p + q & \hs (3 q^2 - 4 p^2 + 2 p q + (2 p - q)\sqrt{4 p^2 -3 q^2})/3 q^2 \hs
& p - q &- 1 \\ \hline
Z             & p - q & (3 q^2 - 4 p^2 - 2 p q + (2 p + q)\sqrt{4 p^2 - 3 q^2})/3 q^2
& p + q &+ 1 \\
\hline U^{\alpha} &  p    &  (4 p^2 -  2 p \sqrt{4 p^2 - 3 q^2})/3 q^2
& - p   & 0  \\ \hline
V^{\alpha} &  q    & ( 3 q^2 - 2 p q + q \sqrt{4 p^2 - 3 q^2})/3 q^2
&   q   &+ 1 \\ \hline\end{array}$$
\caption{Charges of the bifundamental fields
present in the \cY$^{p,q}$ quivers found in \cite{Benvenuti:2004dy}.}\label{charges}
\end{center}\end{table}

\subsection{Chiral building blocks}\label{cbb}
The basic chiral operators correspond to some loops in the quiver that
 have been considered in \cite{Benvenuti:2004dy} in order to give a
 field theoretical computation of the topology of the supersymmetric
 $3$-cycles. Moreover, the analysis of the short loops (R-charge $2$
 chiral ring) has been given in \cite{Benvenuti:2005wi} in order to
 determine the conformal manifold. We will, for the sake of clarity,
 describe nevertheless in detail the various operators also here.

 Let us start from some simple examples of mesonic chiral operators. The simplest chiral single trace operators are of the form\footnote{With the hope of helping in visualize the operators on the quivers, we denote by $Y_c$ the $Y$-fields entering a cubic superpotential term, and by $Y_q$ the $Y$-fields entering a quartic superpotential term.}
 \be\label{exshort}
 tr ( Z \, U \, Y_q \, U ) \;\;\;\; \mbox{or}  \;\;\;\; tr ( U \, V \, Y_c )
 \ee
for 'short' loops of the quivers. For 'long loops' of the quivers of table \ref{examplefig} one finds
 \be\label{exlong1}
 tr ( Z \, U \, V \, U \, V \, U \, Z \, U )
 \ee
 for counter-clockwise loops, and
 \be\label{exlong2}
tr ( Y_q \, U \, Y_q \, U \, Y_c \, Y_c ) \;\;\; \mbox{or} \;\;\;
tr ( Y_q \, U \, Y_q \, Y_c \, Y_c \, U ) \;\;\; \mbox{or} \;\;\;
tr ( Y_q \, U \, Y_q \, Y_c \, U \, Y_c )
 \ee
for clockwise loops. These examples are valid for $Y^{4, 2}$, in
 general there are operators like (\ref{exshort}) of length $3$ and
 $4$, operators like (\ref{exlong1}) of length $2 p$ and
 (\ref{exlong2}) of length $2 p - q$. These three types of operators
 constitute the basic building blocks for any scalar chiral
 operator. Notice that operators corresponding to 'long loops' carry a
 non zero winding number around the quivers; this winding number
 counts the value of the charge associated to the $U(1)$ flavor
 symmetry. Another thing that can be observed immediately is that the
 baryonic charge is always vanishing for any mesonic operator (this
 gives the constraints on the topology of the SUSY $3$-cycles of the
 Sasaki-Einstein manifolds \cite{Benvenuti:2004dy}). As a consequence, we do not consider the baryonic charge in the remaining part of this paper.

\begin{table}[h]\begin{center}
$$\begin{array}{cccc}
\!\!\! {\includegraphics[height=3.7 cm]{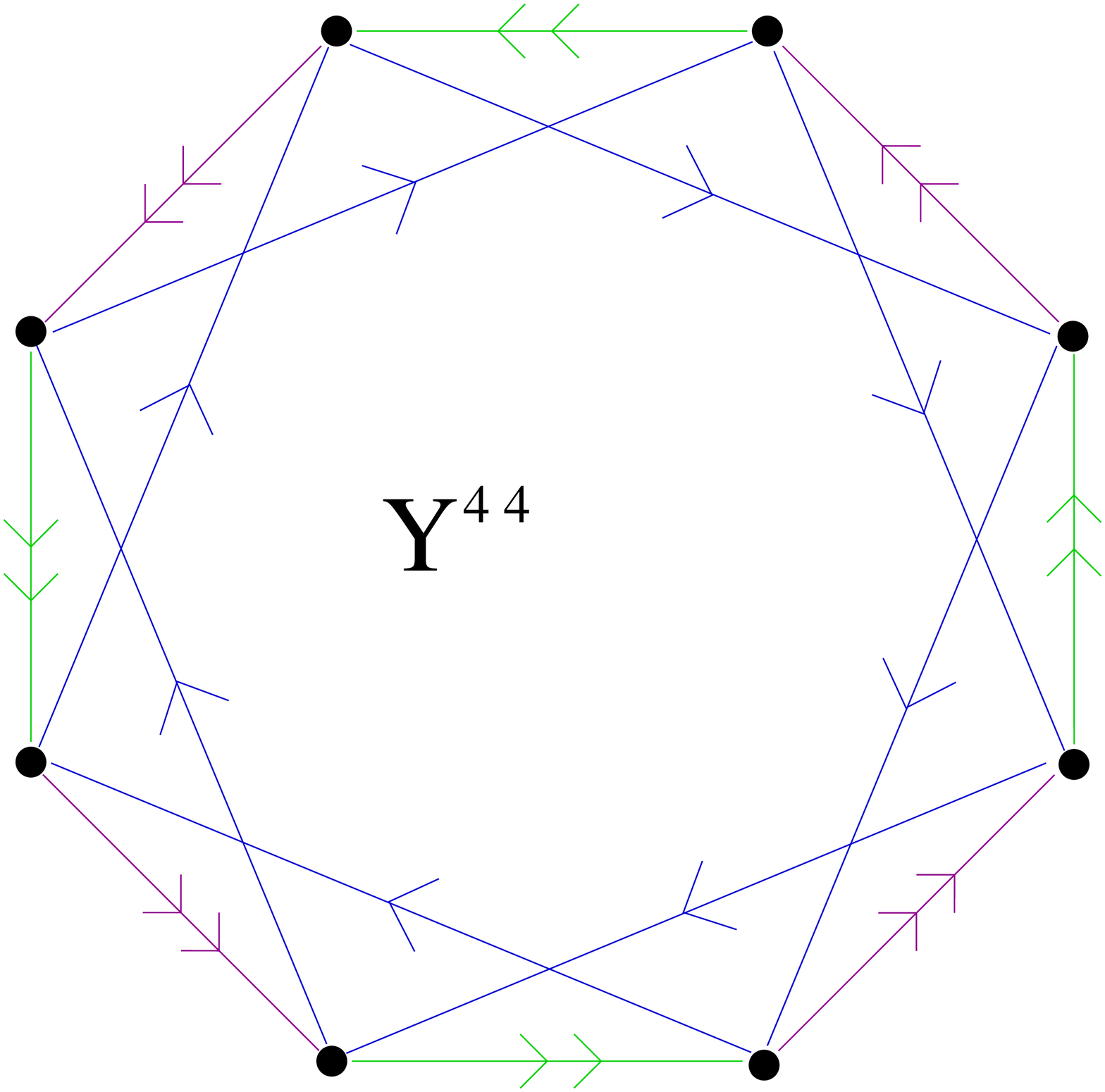}} \;  &
 {\includegraphics[height=3.7 cm]{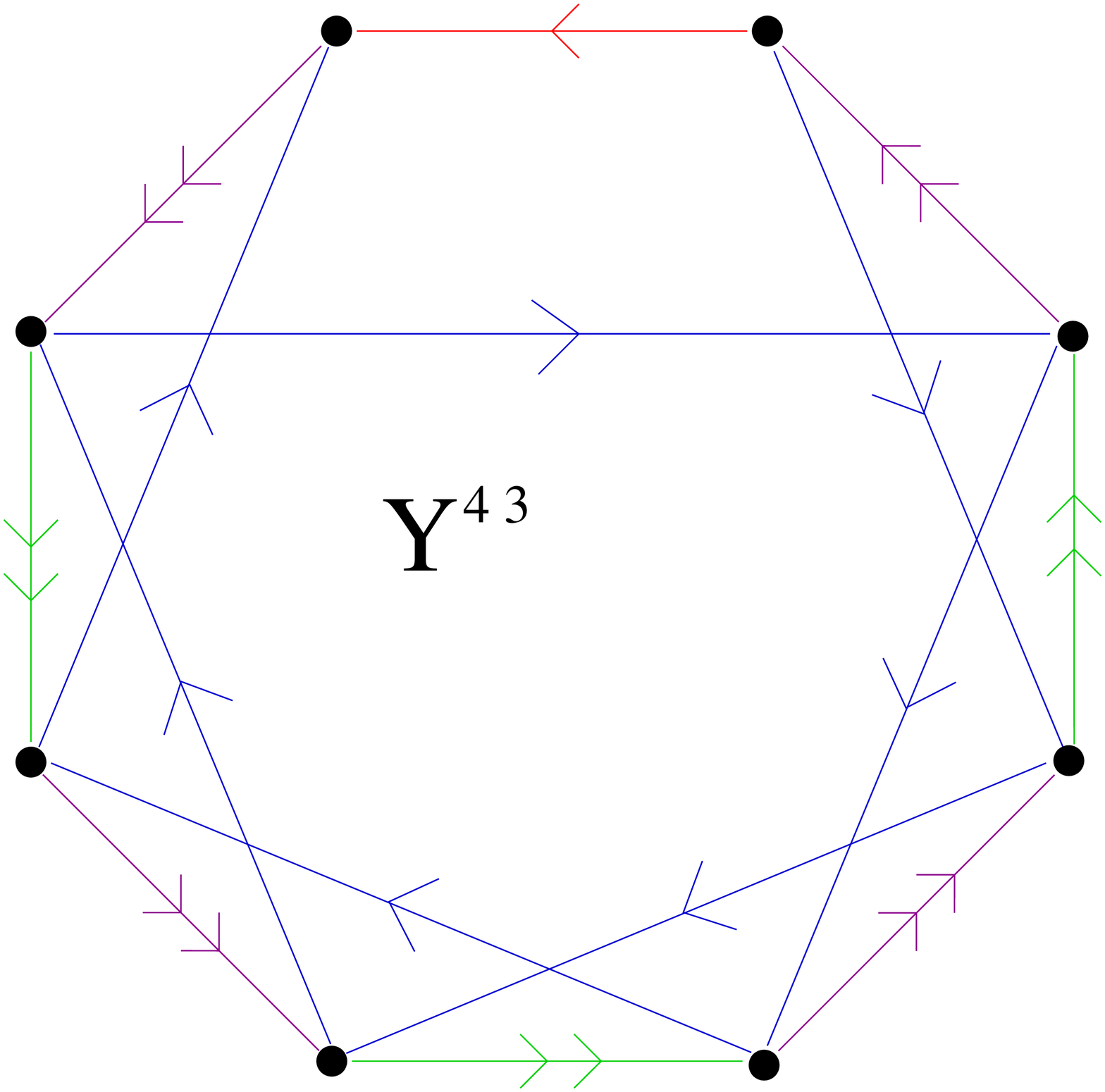}} \;  &
 {\includegraphics[height=3.7 cm]{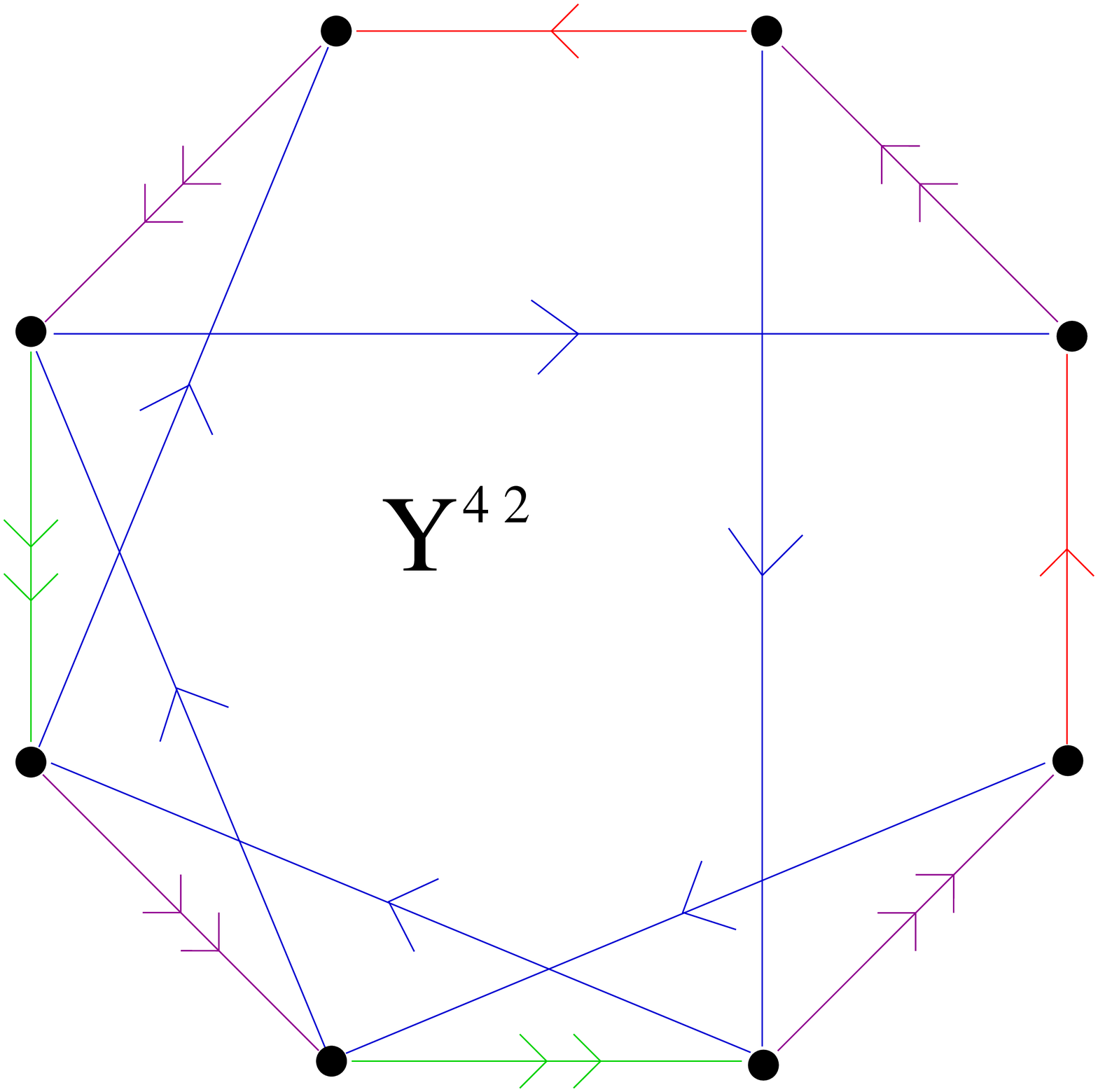}}\;  &
 {\includegraphics[height=3.7 cm]{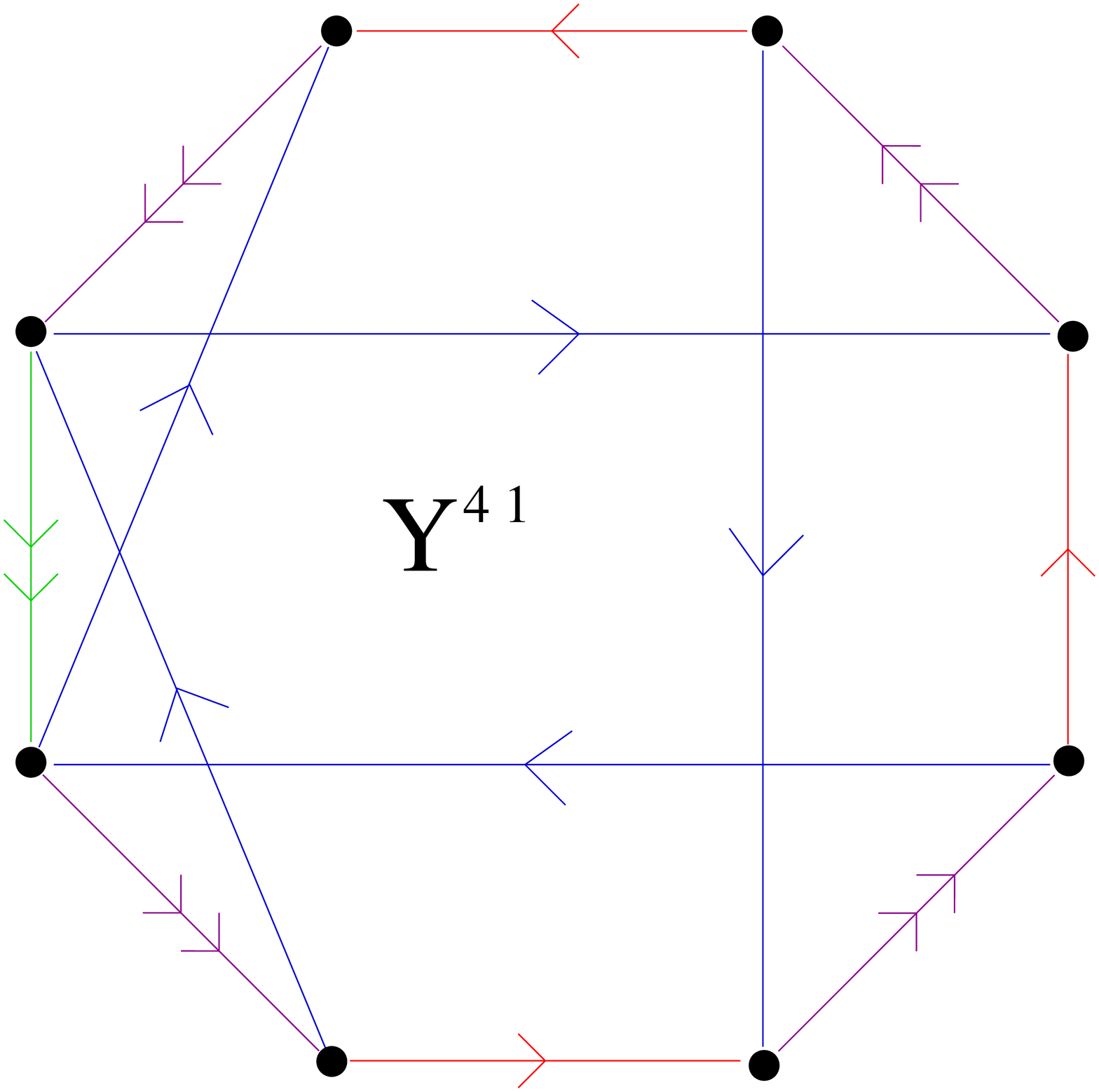}}  \end{array}$$
\caption{Example of the recursive construction of the \cY$^{p, q}$ quivers, as from \cite{Benvenuti:2004dy}.} \label{examplefig}
\end{center}\end{table}

Because of $\mF$-terms relations only a subset of these holomorphic operators are chiral primaries, or BPS, and have dimension $\Delta = 3/2 R$ on the whole IR conformal surface.

Let us consider the short loops, having R--charge $2$. There are $2 p$ such loops: $2 q$ of length $3$ and $p - q$ of length $4$. Moreover, since the fields $U$ and $V$ transform in the spin--$1/2$ of the global $SU(2)$, all the short loops are in the $1/2 \otimes 1/2 = 0 \oplus 1$. We are thus dealing with $4 ( p + q )$ operators. The $\mF$-term relations will imply that only $3$ of them are chiral. The explicit superpotential \cite{Benvenuti:2004dy} is
\beq\label{WYpq}
W = \sum_{i = 1}^q \epsilon_{\a \b}
( U_i^\a V_{i}^\b Y_{2i - 1} +  V_i^\a U_{i+1}^\b Y_{2i} )
    + \sum_{j = q + 1}^p \epsilon_{\a \b} Z_j U_{j+1}^\a Y_{2j - 1} U_j^\b\;.
\eeq
It is important to remember that this writing is schematic, the precise coefficients multiplying every $SU(2)$-invariant term depend on the position in the conformal surface, similarly to the gauge couplings.
The equations of motion of the $Y$-fields
 \bea
 U^1_i V^2_i & =  & U^2_i V^1_i \\
 V^1_i U^2_{i+1} & =  & V^2_i U^1_{i+1} \\
 U^1_j Z_j U^2_{j+1} & = & U^2_j Z_j U^1_{j+1}
 \eea
immediately say that the spin--$0$ parts are zero in the chiral ring. The equations of motion for the 'external' $U$, $V$, $Z$ fields enable to 'move' the short loops around the quiver. All these short loops are thus equal in the chiral ring. The superconformal BPS operator is a symmetrization over the quiver of all these short operators:
\beq\label{shortexplicit}
\cS^I = \sum_{i = 1}^q \sigma^I_{\a \b}
( U_i^\a V_{i}^\b Y_{2i - 1} +  V_i^\a U_{i+1}^\b Y_{2i} )
    + \sum_{j = q + 1}^p \sigma^I_{\a \b} Z_j U_{j+1}^\a Y_{2j - 1} U_j^\b\;,
\eeq
where $\sigma^I$ are the $3$ Pauli matrices. Also in (\ref{shortexplicit}) the precise coefficients in front of every $SU(2)$-covariant term depend on the position in the conformal surface.
In conclusion there are only $3$ operators, $\cS^\pm$ and $\cS^0$, with R--charge $2$ and scaling dimension $\Delta = 3$ over the whole conformal surface, and they transform in the spin--$1$ representation of $SU(2)$. Note that the $Q_F$ charge of $\cS$ is $0$. The chiral operator with vanishing spin--z, $\cS^0$,  lies in the chiral ring for any toric superconformal quiver and drives the exactly marginal deformation called $\b$--deformation \cite{Benvenuti:2005wi}. For more general toric quivers, without an ``accidental'' $SU(2)$ global symmetry, BPS building blocks with non vanishing $U(1) \times U(1)$ flavor charge, like $\cS^{\pm}$, are not short: their analog are similar to the 'long' chiral operators we are going to study now.

Considering the winding operators we distinguish between clockwise and counter-clockwise loops. The length-$2p$ loop (counter-clockwise, of the form (\ref{exlong1})) is made of $p$ $U$-fields, $q$ $V$-fields and $p-q$ $Z$-fields. The set of operators (in total $2^{p+q}$) corresponding to this loop transform in the $SU(2)$-representation with spin
 \be
  \left( \otimes^p \frac{1}{2} \right) \otimes\left( \otimes^q \frac{1}{2} \right) =
   \frac{p + q}{2} \oplus \frac{p + q - 2}{2} \oplus \dots
 \ee
The $\dots$ represent lower dimensional $SU(2)$-representations. All the non-maximal $SU(2)$-representations in the chiral ring vanish, due to the $Y$-fields $\mF$-term relations. We denote this operator $\cL_+$. One thus finds that it carries spin $\frac{p + q}{2}$ and, from Table \ref{charges} a non vanishing positive $U(1)_F$ charge.

The other type of winding operators, clockwise loops of the form (\ref{exlong1}), are a little bit more difficult to visualize. They have length $2 p - q$ and are made of $p$ $Y$--fields and $p - q$ $U$--fields (including $U$- or $Z$-fields in such a loop is equivalent to multiply by an $\cS$ operators, and would not be a building block). The $Z$- and $Y$-fields $\mF$-term relations imply that the $SU(2)$ indices have to be completely symmetrized, i.e. the BPS operators transform in the spin-$\frac{p-q}{2}$ representation. On the other hand, $V$-field $\mF$-terms enable one to move the position of the various $U$- and $Y_c$-fields present in the operators of the form (\ref{exlong1}). This implies that there is only one BPS clockwise loop, with $J= \frac{p-q}{2}$, that we call $\cL_-$.\footnote{In the case of \cY$^{p,p}$ there are two different $\cL_{-}$ operators. This fact does not lead to a strong enhancement of the chiral ring, since in this case $J=0$, in particular it is not possible to use this two operators to construct long operators dual to non point-like semiclassical strings.}

\begin{table}[!h]
\begin{center}
$$\begin{array}{|c|c|c|c|}  \hline
\hs\mathrm{Meson}\hs& \hs \mathrm{spin} \hs J \hs &   Q_R    &\hs Q_F \hs \\ \hline\hline
 \cS          &     1             &           2              &  0         \\\hline
 \cL_+        &\frac{p + q}{2}    &\hs p + q - \frac{1}{3 \ell} \hs& + p        \\\hline
 \cL_-        &\frac{p - q}{2}    & p - q + \frac{1}{3 \ell} & - p        \\ \hline
\end{array}$$
\caption{Charge assignments for the three basic mesonic fields. Notice that $2 J - Q_R$ is proportional to $Q_F$ for all the operators.}
\label{mesoncharges}
\end{center}
\end{table}

The R--charge of the long loops is computed using Table \ref{charges}
 \be
 Q_R[\cL_{\pm}] = p \pm (p( 2 p - \sqrt{4 p^2 - 3 q^2} ))/ 3 q =
 p \pm (q - \frac{1}{3 \ell})\;,
 \ee
where in the last equality the relation (\ref{defell}) has been used. The final results are summarized in Table \ref{mesoncharges}.

The natural way to think of the mesonic operator is in terms of the
quiver diagram drawn on a two--torus, as suggested by the connection
between dimers and toric geometry \cite{Hanany:2005ve, Franco:2005rj} . The point is that any toric quiver can be drawn on a torus in such a way to provide a polygonalization of the torus. The quiver diagram is precisely the dual diagram of the dimer model. Each face is surrounded by bifundamental fields going either in the clockwise or in the counterclockwise direction, and precisely corresponds to a superpotential term. For the toric phases of \cY$^{p,q}$ quivers there are only cubic and quartic superpotential terms, so all the faces are triangles or squares. The torus has one 'short' homology cycle and one 'long' homology cycle.

In this picture, the generators of the \cY$^{p, q}$ chiral ring are as follows:
\begin{itemize}
\item $\cS^0$ is the only chiral operator that does not wind around any homology cycle of the torus.
\item $\cS^{\pm}$ wind around the short homology cycle, in opposite directions.
\item $\cL_{\pm}$ wind around the long homology cycle. The value of the $z$--spin counts the winding number around the short homology cycle.
\end{itemize}

A general fact of toric quivers we find is that the values of the two commuting $U(1)_F$ charges (that are always present) are counted precisely by the two winding numbers of the operator. This is actually valid for any mesonic operator, not just BPS ones.

Of course the generators of the chiral ring we found satisfy various non linear relations. Studying these relations, it should be possible to reconstruct the algebro-geometric description of the \cY$^{p, q}$ Calabi-Yau cones. Instead of doing this, we will reconstruct the transverse geometry through the analysis of semiclassical holomorphic operators, in sec. \ref{spin chain action}. This will give also information about the metric.

\subsection{The full mesonic chiral ring}\label{mcr}
We now want to consider 'multiloop' operators. First of all we see what happens multiplying two $\cS$. An operator like
 \be
 tr(U_i V_i Y_{2i+2} U_i V_i Y_{2i+2}) \sim  tr(U_i V_i Y_{2i+2} U_i Y_{2i+3} V_{i+1})
 \ee
transforms in the $\otimes^4 1/2 = 2 \oplus \dots$ and can be seen
simply as the product of two short mesons $\cS$. It is easy to
convince that this is general: any chiral operator that does not
wind all the way around the quiver is of the form $\cS^n$ and
transforms in the spin--$n$ of $SU(2)$. In other words, BPS operators do not carry a position on the quiver, and always transform in the maximal possible $SU(2)$-irrep.

Consider now the product of $\cL_+$ and $\cL_-$. This product,which winds $0$ times around the quiver, can be expressed in terms of the $\cS$ operators.
 \be
 \cL_+ \cL_- \sim tr( \dots U V U \; Y Y U \dots ) \sim
 tr( \dots U V Y \; U Y U \dots )
 \ee
where we used the equation of motion for $V$. We thus see that the resulting operator is the product of various $\cS$s. More precisely
 \be
 \cL_+ \cL_- \sim \cS^{p}
 \ee
Note that this relation is consistent with the charge assignments of Table \ref{mesoncharges}.

We are now in the position of giving the classification of the mesonic chiral BPS operators of the $Y^{p, q}$ quivers. A general operator $\cO$ can be seen as the product of $\cS$ and $\cL$:
 \be
 \cO_{s, \, l} = \cS^s \cL^l
 \ee
Where we denote $\cL = \cL_+$ and $ \cL^{-1} = \cL_-$. $s$ is a
non-negative integer, while the integer $l$ can be positive of
negative. The R--charge of $\cO_{s, l}$ is given by
 \be\label{Rgene}
 Q_R[\cO_{s, \, l}] = 2 s + p |l| + l \left(q - \frac{1}{3 \ell} \right)
 \ee
while the flavor charge
 \be
 Q_F[\cO_{s, \, l}] =  p \, l
 \ee
Finally, $\cO_{s, \, l}$ transforms in the irreducible $SU(2)$-representation with spin J
 \be
 J[\cO_{s, \, l}] =  s + |l| \frac{p}{2} + l \frac{q}{2}
 \ee

Again, the precise form of these operators can be obtained by a complete symmetrization over the quiver (imposed by $U$- and $V$-fields $\mF$-terms) and over the $SU(2)$ indices (imposed by $Y$- and $Z$-fields $\mF$-terms). A complete symmetrization over the trace is also to be performed.

\subsection{BPS geodesics from the quivers}

 The operators corresponding to point-like strings moving along a null BPS geodesic are chiral primaries and therefore
should be among the ones we just described. In this section we make the mapping precise and compare with the results of
Sec. \ref{massless}.

Before doing so note that, heuristically, we can understand that these operators correspond
to point-like strings because, due to the complete symmetrization imposed by $\mF$-term relations, the three values of
the $U(1)^3$ charges are constant along the operator, for long
operators and in a sense made precise later  when we
study a coherent state representation of the operator (see eqs. (\ref{long1})
and (\ref{long2})). There we also see that we can get non-BPS operators by tuning
continuously, along the operator, the ratio $l/s$, corresponding to the value of the $U(1)_F$ charge and the $z$-component
of the $SU(2)$, corresponding to the difference between the number of $A^1$-fields and the number of $A^2$-fields
($A^\a$ stands for $U^\a$ or $V^\a$.) The R-charge is determined in term of $l$ and $s$ by the relation \ref{Rgene}.

 Going back to our main problem in this section, the first task is to reobtain, from the field theory,
the quantities $y_{1,2}$ and $\ell$ (defined in (\ref{yla})) that play an important role in the supergravity background.

We start by writing the charges of a chiral operator made out of $n_+$ operators $\cL_+$, $n_-$ $\cL_-$'s and $s$ operators $\cS$ composed to maximum $SU(2)$ spin $J$. The result is
\beqa
Q_F &=& pn_+ - pn_- = p n_{\alpha} \\
J\; &=& n_+ \frac{p+q}{2} +n_-\frac{p-q}{2}+s = \half ( p\bar{n} +2s) + \half q n_{\alpha}\label{chch0}\\
Q_R &=& n_+ \left(p+q-\frac{1}{3\ell}\right) + n_-\left(p-q+\frac{1}{3\ell}\right) + 2s=
(p\bar{n} + 2s) + \left(q-\frac{1}{3\ell}\right)n_{\alpha}\phantom{aaaa}\label{chch}
\eeqa
where we introduced $\bar{n}=n_++n_-$ and $n_{\alpha}=n_+-n_-$. We see that we can use $n_{\alpha}$ instead
of $Q_F$. Furthermore, $\bar{n}$ and $s$ appear only in the combination $p\bar{n} + 2s$ which follows from
the fact that actually $\cL_+\cL_-\sim \cS^p$ in the chiral ring. This means that there are only two independent
numbers and therefore from (\ref{chch0}) and (\ref{chch}) a relation between the charges follows
\beq
Q_R - 2 J =
-\frac{1}{3\ell} n_{\alpha}
\eeq
 So, $\ell^{-1}$ has appeared as a natural unit for $U(1)_F$ charge. We can define two new variables:
\beq
 \Pa = \frac{n_{\alpha}}{\ell} , \ \ \ \ \mbox{and} \ \ \ \  y_0 =
 -\frac{\Pa}{3Q_R}
\label{Pftdef}
\eeq
 Therefore, in the field theory, $y_0$ is the relation between $U(1)_F$ and R-charges for a
given operator. The range of $y_0$ is determined by noticing that its minimum and maximum values
correspond to $\cL_+$ and $\cL_-$ respectively, namely for $n_-=s=0$ and $n_+=s=0$:
\beqa
y_0(\cL_+) &=& -\frac{n_{\alpha}(\cL_+)}{3\ell Q_R(\cL_+)} = y_1 \\
y_0(\cL_-) &=& -\frac{n_{\alpha}(\cL_-)}{3\ell Q_R(\cL_-)} = y_2
\eeqa
where we used the $Q_R$ charges of $\cL_+$ and $\cL_-$ from table \ref{mesoncharges}.
 In this way we recover, from the field theory, that
\beq \label{ylimits}
y_1\le y\le y_2
\eeq
 If we rewrite now the charges (\ref{chch}) in terms of $y_0$ we get perfect agreement with (\ref{PaJ})
and (\ref{DQ}):
\beq
 J = \half (1-y_0) Q_R, \ \ \ \Pa = -3y_0 Q_R
\label{PaJft}
\eeq
 We have therefore identified the chiral operators with massless geodesics. Particular examples
are: an operator made out only of $\cL_+$ with the geodesic at $y_0=y_1$, one made out only of $\cL_-$ with
the geodesic at $y_0=y_2$ and one made out of equal number of $\cL_+$ and $\cL_-$ with the geodesic at $y_0=0$.

\section{The field theory at $\lambda\to0$}\label{lambda0}

 Since BPS operators are protected, the matching between chiral primary operators and massless geodesics is
valid at generic points on the conformal surface of the \cY$^{p, q}$ quivers. The $SU(2)$ invariant points
on the conformal manifolds are parametrized by two complex parameters\footnote{In \cite{Benvenuti:2005wi}
a description of the full conformal manifold is given.}. On
the string side  \cite{kw,Herzog:2004tr, Benvenuti:2004wx}, these parameters are the complex
dilaton ($g_s$) and the vev of the $B$-fields (RR and NSNS) on the $2$-cycle present in
the manifold $S^2 \times S^3$.

 On the string side we further considered a set of string states described by the effective action
of sec. \ref{reducedaction}. Since these states are not BPS the effective action is valid only in the
regime $g_s\ll 1$ (to ignore string loops) and small curvature, namely $\lambda = (R/l_s)^{\frac{1}{4}} =g_s N \gg 1$.

 In the case of \N{4} this effective action can be compared to a similar action derived from the field theory
in the opposite regime $\lambda \ll 1$ (which can also be interpreted as taking $g_s$ to zero keeping $N$ fixed and large).
 In our case, since the effective action is proportional to $\lambda$, a naive extrapolation to $\lambda$ small suggests
that, in that regime, the result might be interpreted as a small perturbation around a point with $\lambda=0$.
 If such a point exists (actually should be a line) it is special since all the semiclassical operators described
by the reduced action would satisfy $\Delta = 3/2 Q_R$.

 In the case of \N{4} SYM (and orbifolds thereof), this point is the free theory. In the case of the conifold
\cite{Benvenuti:2005wi} there is a line of conformal fixed points with vanishing superpotential which is part
of the conformal manifold. Having $\mW = 0$ implies that \emph{all} chiral operators are chiral primaries, or
BPS, operators; in other words the chiral ring is much bigger on this line, depicted in fig. \ref{flow}.
This means that this particular line should be identified with the $\lambda=0$ point as suggested
in \cite{Gomis:2002km} where the BMN limit of the conifold was studied. Notice that the gauge couplings are not zero
which precludes doing standard perturbation theory. Rather, one should do conformal perturbation theory on a
marginal perturbation given by the quartic superpotential.
\FIGURE{\epsfig{file=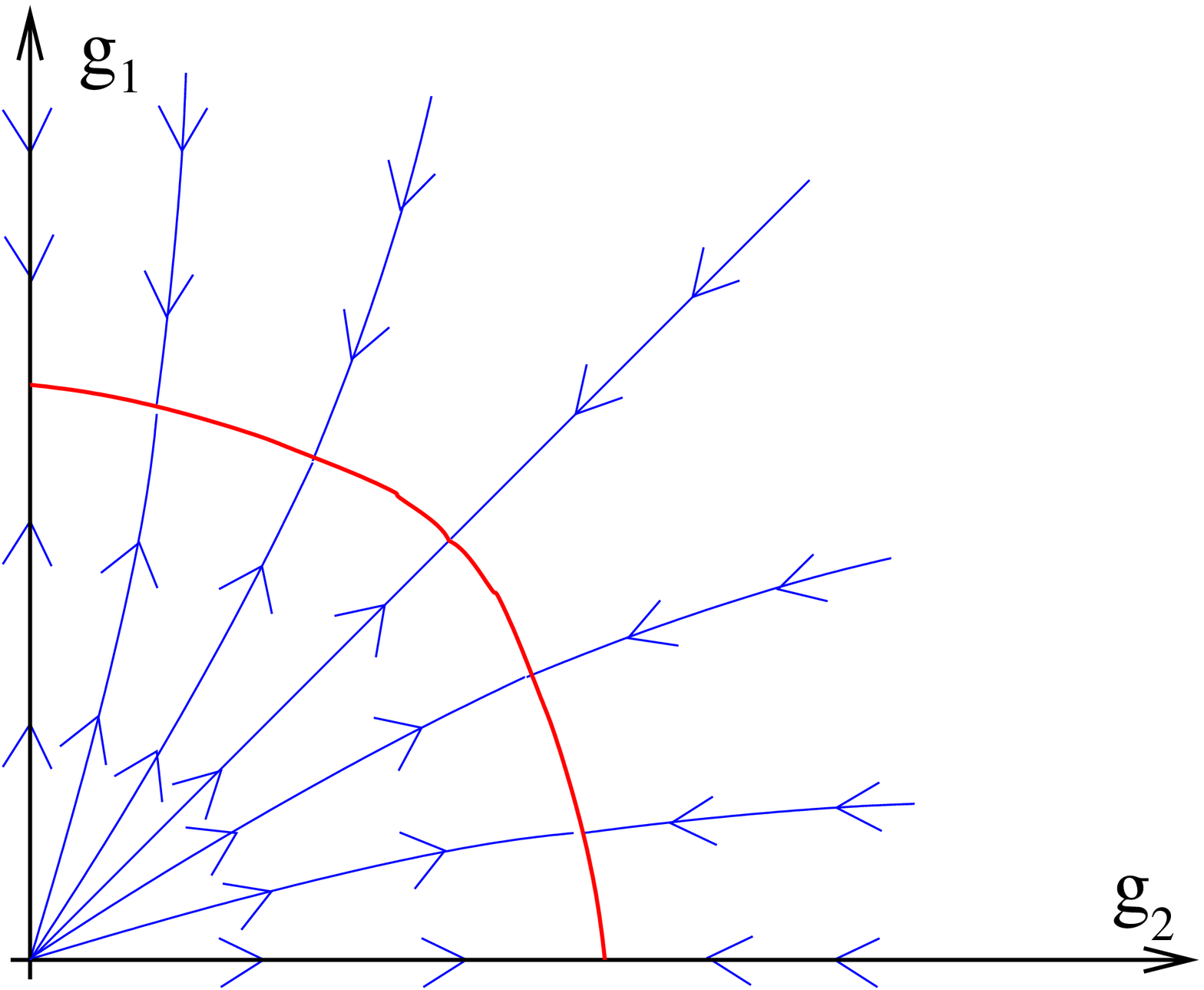, height=3cm}
\caption{Space of couplings for the conifold field theory with no superpotential. The conformal manifold is indicated.}
\label{flow} }
 As a last preliminary example, for the \cY$^{2,0}$ quiver in the ``single impurities phase'', this $\mW = 0$
conformal line has an enhanced chiral ring and a global symmetry $SU(2)^4$. What is interesting here is that
one can Seiberg dualize \cite{Seiberg:1994pq} this phase with $\mW = 0$, obtaining a special line on the
``double impurity phase'' \cite{Benvenuti:2004wx}; in this dual description there are the standard
Seiberg $\mathcal{M} q \tilde q$ superpotential terms, so the superpotential does not vanish. We conclude
that $\lambda = 0$ on the string side is not, in general, equivalent to $\mW = 0$ on the gauge side.

 We can now consider a generic \cY$^{p,q}$ quiver. In this case the superpotential has cubic and quartic terms.
Since there are several terms in the superpotential we do not expect to be able to cancel all of them since we
can only vary two parameters on the conformal surface. If only gauge couplings are
turned on, the conformal dimension decreases and we can never get to the conformal manifold where for example
the $V$-fields have R-charge larger than $2/3$\footnote{The presence of fields with $Q_R \geq 2/3$ follows,
directly, from the presence of nodes with $N_F = 3 N_C$.}.  We conclude that, at least for the toric phases,
there is no point on the conformal manifold with vanishing superpotential.

 In fact this problem was studied in \cite{Benvenuti:2004dw}, where it is pointed out that, if a IR conformal
manifold exists\footnote{This is not the case for a generic \N{1} quiver theory.}, it is sufficient that
\emph{one} gauge group is asymptotically free in order to be able to flow from the free theory to the conformal
manifold but, in order to reach this point, the superpotential couplings are crucial. For the \cY$^{p,q}$ models,
the flow has been qualitatively described in \cite{Benvenuti:2005wi}: one flows at first the $N_F = 2 N_C$ nodes,
then some cubic superpotential terms and so on.

 We can actually go further by using an idea of Kutasov
 \cite{Kutasov:2003ux}, that can be thought of as extending the $a$-maximization
procedure of Intriligator and Wecht \cite{intriligator03} away from the conformal manifold. In our case this
consists in introducing $2p$ Lagrange multipliers $\lambda_{i=1\ldots2p}$, one for each gauge group, and
$\mu_{k=1\ldots p+q}$, one for each term in the superpotential. We can define $a$ as:
\beq
a = 3 \tr (R-1)^3 - \tr (R-1) - \sum_i \lambda_i \left(\tr( \epsilon_i (R-1)) +2 \right) + \sum_k \mu_k (\tr(\nu_k R)-2)
\eeq
where $\epsilon_i$ is a diagonal matrix in the space of fields which is $1$ or $0$ if the corresponding field
is charged or not with respect to the $i$-gauge group. The same for $\nu_k$ which is $1$ or $0$ if the field
appears or not in the $k$-th term in the superpotential. If we maximize $a$ with respect to the
R-charges they become functions of the Lagrange multipliers $(\lambda_i,\mu_k)$. It was further argued
in \cite{Kutasov:2003ux} that the Lagrange multipliers can be used to parameterize the space of couplings.
Although the relation between the couplings and the multipliers is
still somewhat conjectural, one thing is clear: a coupling is zero if
and only if the corresponding Lagrange multipliers is zero, since the
corresponding constraint has not to be imposed. We want to see now if, on the conformal surface, we can make some of the Lagrange
multipliers to vanish. Since we are on the conformal surfaces we have to impose anomaly cancellation conditions.
After that one can see, working in specific examples, that one can put some quartic terms in the superpotential
to zero. (Here we are also using that all Lagrange multipliers are positive
in the physical region of the couplings).


 We can now understand the superpotential corresponding to $\lambda=0$. Decreasing $\lambda$, all the couplings
(taken in their absolute value) decrease. At some point a coupling becomes zero. This coupling has thus to
correspond to a quartic term in the superpotential.

 Having a vanishing superpotential coupling generates a change in the chiral ring of the theory but it is not
obvious that the holomorphic long operators corresponding to the semiclassical strings of sec. \ref{reducedaction}
become protected. To see that we focus on an example, \cY$^{4,3}$, which should clarify the general structure.

Consider the operator $\cL_+$ in the \cY$^{4,3}$ quiver.
\beq
\cL_+= tr ( Z U V U V U V U)
\eeq

\FIGURE{\epsfig{file=y43.eps, height=4.5cm}
\caption{Quiver diagram corresponding to $\cY^{4,3}$. }
\label{y43} }

 To be chiral primary, namely not a descendant, $\cL_+$ has to  transform in the spin-$7/2$
representation of $SU(2)$, as follows from using the $\mF$-term relation of the $Y$-field entering
in the cubic terms.  If we multiply two $\cL_+$ operators, for $\mW_4 \neq 0$, we saw in sec. \ref{mcr}
that we get an operator that transform in the spin-$7$ representation (with $15$ states), of the form
\beq
 tr ( Z \, U V U V U V U \, Z \, U V U V U V U )
\eeq
If $\mW_4 = 0$, however, there are more than $15$ states. The reason is that one cannot use the $\mF$-term relations
coming form the quartic terms to move the $SU(2)$ spins from the first $U V U V U V U$ block to the second $U V U V U V U$
block. More generally, operators of the form $(\cL_+)^n$ contain a number of states that grows as $8^n$ if $\mW_4 = 0$ and
as $8 n$ if $\mW_4 \neq 0$. It is also possible to see that the spin-$1$ $\cS$ operator gets enhanced to
spin-$(1 \oplus 0)$. This $\cS$ operator with $J=0$ generates an exactly marginal deformation, which is precisely
the conformal line parameterized by $g_s$.

 What happens is thus that, at $\mW_4 = 0$, the chiral ring is much larger which leads us to identify this point
with $\lambda=0$ although it is not true that all chiral operators are in the chiral ring.

We consider now semiclassical operators at the $\mW_4 = 0$ point. Let us focus for simplicity on operators of the
form $(\cL_+)^n$. These correspond to semiclassical strings moving only on the round two-sphere, satisfying
$y (\sigma , \tau) = y_1$. A class of operators of the form $(\cL_+)^n$ is as follows:
\beq\label{yofsigma}
tr \left( \prod_{i=1}^{n} R(\theta_i,\phi_i) \cL_+ \right)
\eeq
 where $R(\theta_i,\phi_i)$ is an $SU(2)$ rotation applied to $\cL_+$ (which has maximum $z$-spin $P_\phi$).
Taking $n$ to be large, and the angles $(\theta_i, \phi_i)$ which parameterize the rotation to vary smoothly with $i$,
we see that we are constructing a semiclassical string extended along the S$^2$ sphere parameterized by $(\theta,\phi)$.
 This is similar to what happens in the $SU(2)$ sector of \N{4} operators \cite{kru}. Reconstructing the directions
$y$ and $\beta$ is more involved and can be recovered from the results
 the next section (naively, the absence of the $\mF$-term relations
 coming from $\mW_4$ implies that one cannot exchange $\cL_+$  with $\cL_-$). We just emphasize that the important point here is that a generic operator like (\ref{yofsigma}), at $\mW_4 = 0$, is BPS, and satisfies the relation $\Delta = 3/2 Q_R$.



\section{Effective action for the spin chain} \label{spin chain action}

 In the previous sections we studied conformal primaries and compared them to the massless geodesics
in the metric. As we just discussed, going further is difficult since the theories are strongly coupled
and we cannot use a perturbative expansion to compute anomalous dimensions. In principle, as argued in the
previous section, we should use conformal perturbation theory around a conformal point where some terms in
the superpotential vanish. Instead of doing that, to simplify the problem, we are going to consider all terms
of the superpotential on equal footing and extract a simple spin chain model that captures the generic
features of the operator mixing that the superpotential produces. Even then we are going to simplify the problem
further. From the point of view of the resulting spin chain what we are doing is trying to obtain the correct
long distance physics so we expect that the microscopic details should not be important.

 Using coherent states we obtain an effective action for the spin chain which is similar to the one we
derived from string theory, namely eq. (\ref{redaction}) albeit with a different function $p(y)$.

 We analyze first the case of \cY$^{3,2}$ which should make clear the generic case we discuss afterwards.

\subsection{Long paths in the \cY$^{3,2}$ quiver}

 The quiver corresponding to \cY$^{3,2}$ is depicted in figure \ref{figY32}. Gauge invariant operators
correspond to closed paths in the quiver. An example is the outer counterclockwise loop that we called
$\cL_+$. Other important example is the operator $\cL_-$. It is a linear combination (with equal coefficients)
of the three paths depicted in the figure and can be written as:
\beq
 \cL_-^{(3,2)} = \frac{1}{\sqrt{3}}\left[ \tr(Y_q UYY) + \tr(Y_q YUY) + \tr(Y_q YYU)\right]
\eeq
 Here, for clarity, we denoted as $Y_q$ the operator $Y$ that appears in the quartic
superpotential term $\tr(U Y_q UZ)$. We see that $\cL_-$ is a mix of three operators where the
operator $U$ moves between three possible positions among the $Y$'s. This mixing comes from the
cubic vertices of the superpotential as can be seen in the example of figure \ref{figmoves}.
 The mixing matrix induced by these vertices is proportional to
\beq
 H = h \left( \begin{array}{ccc}-1 & 1 & 0 \\ 1 & -2 & 1 \\ 0 & 1 & -1 \end{array}\right)
\eeq
 The off-diagonal terms correspond to the mixing. The diagonal terms come also from the superpotential.
They have opposite sign due to the relative sign between different terms in the superpotential and one is
double of the others since, for that state,  the $U$ has two neighboring $Y$'s. Instead, when the $U$ is
between $Y_q$ and $Y$ only the $Y$ counts since there is no term in the superpotential involving $U$ and
$Y_q$ (and no $Z$). We wrote the mixing matrix as $H$ since
one can think of it as a Hamiltonian whose eigenvalues are the conformal dimensions. In this case we find the
eigenvector $\frac{1}{\sqrt{3}}(1,1,1)$ with eigenvalue zero which is precisely $\cL_-$.
 The constant $h$ denotes the superpotential couplings (and other
 factors appearing in the computation) and therefore cannot be taken
 to be small  in general. This implies that to obtain the correct
spin chain  Hamiltonian one should use non perturbative techniques that sum all
 the diagrams. What is clear is that $\frac{1}{\sqrt{3}}(1,1,1)$ is
 always a ground state, since it is a protected operator.

\FIGURE{\epsfig{file=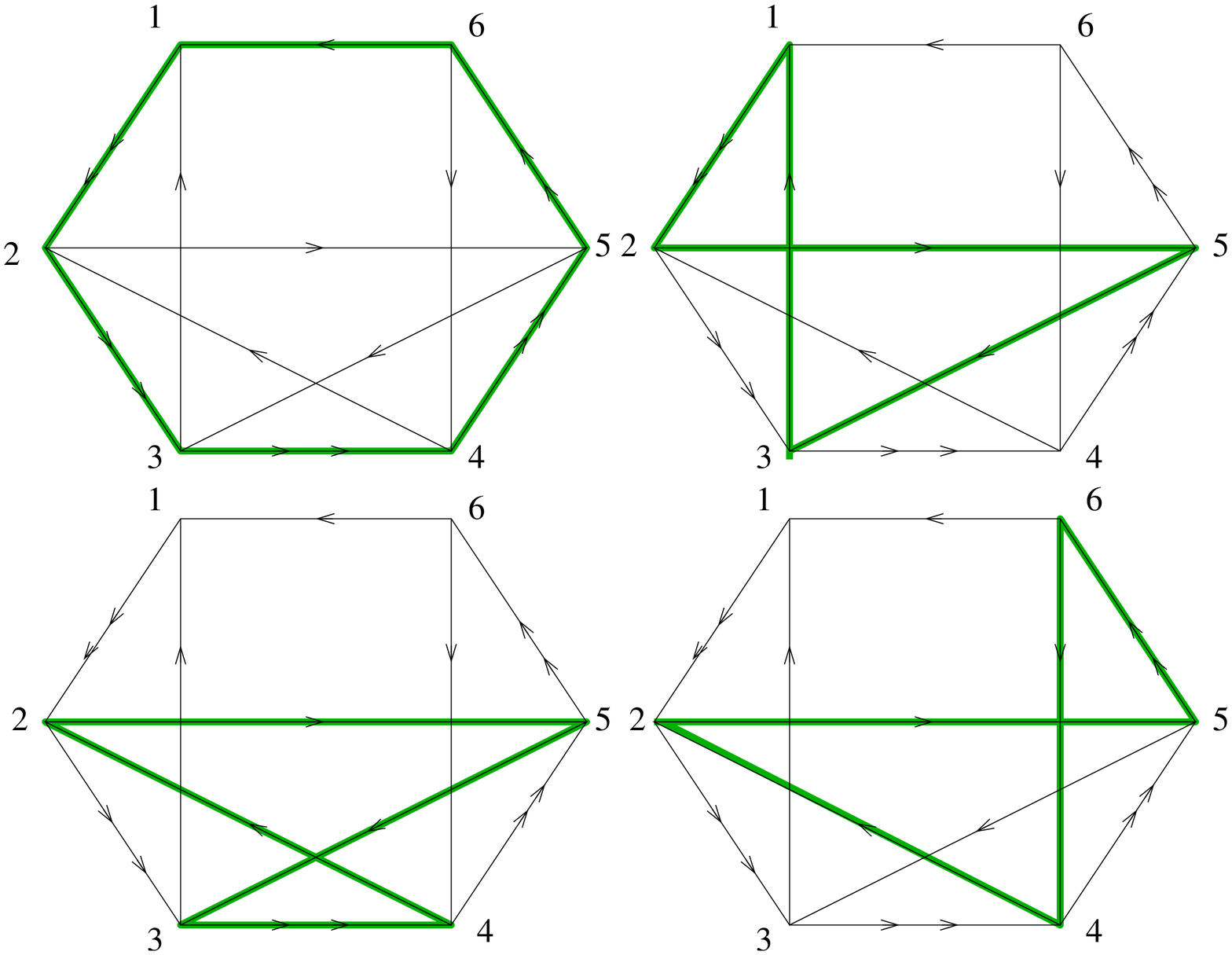, height=10cm}
\caption{Quiver diagram corresponding to $\cY^{3,2}$. We show, on the top left, the path  $\cL_+$. The other
three paths are those whose symmetric linear combination is $\cL_-$.  }
\label{figY32} }

Now we should investigate what happens for more generic operators,
namely to all possible closed loops in the quiver. These loops form a basis in a Hilbert space. In such space
we can define a Hamiltonian that converts a given path in a linear combination of all paths that can be
obtained from it by using the ``moves'' of the type described in figure \ref{figmoves}.

\FIGURE{\epsfig{file=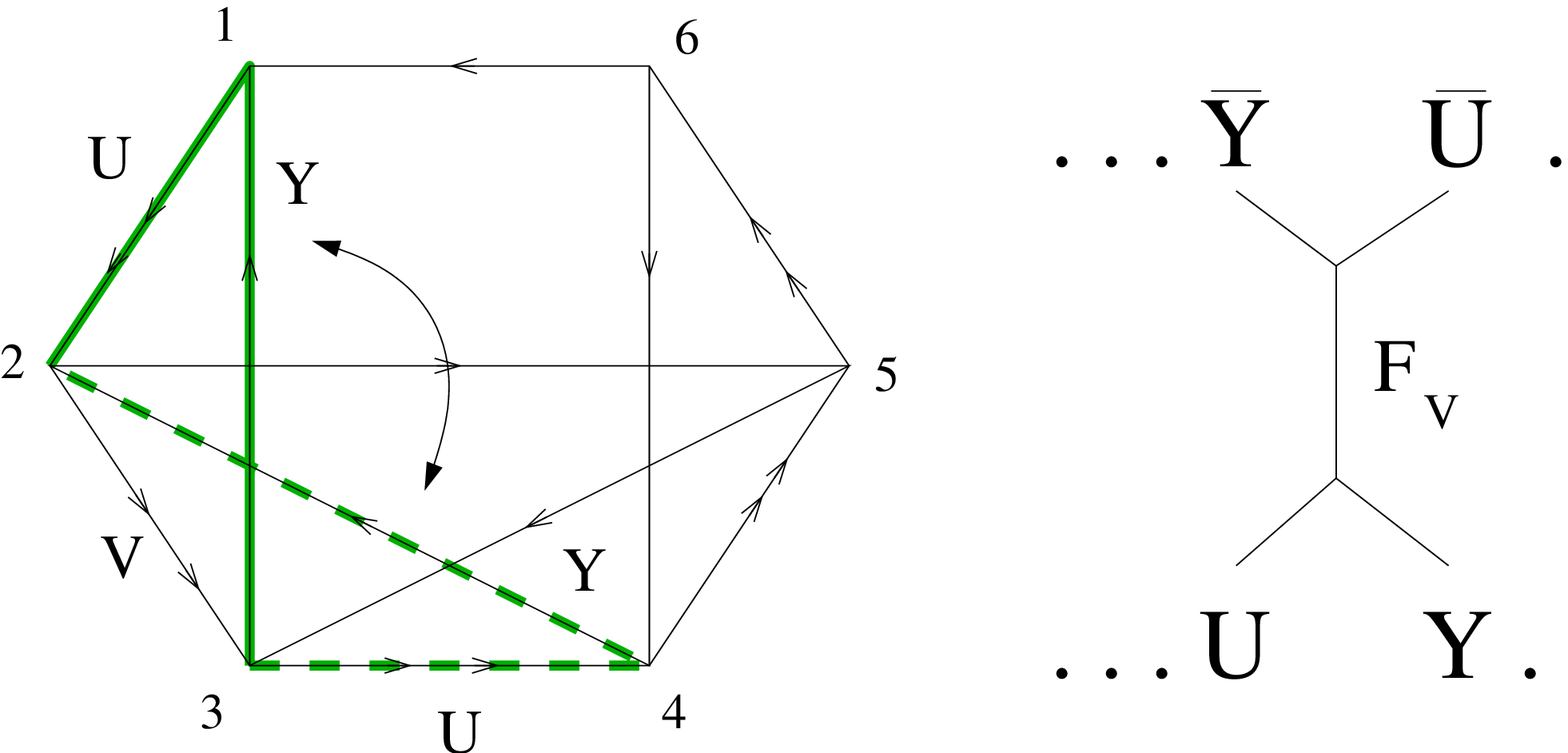, height=6cm}
\caption{The superpotential generates mixing among the chiral operators, namely the closed paths in the quiver.
Diagrams as the one shown on the right give mixing between paths that differ by the ``move'' shown on the left:
$324\leftrightarrow 342$. Here $F_V$ denotes the $F$ component of the chiral field $V$.
The other moves are: $423\leftrightarrow 453$, $534\leftrightarrow 564$, $6125\leftrightarrow 645$,
$231\leftrightarrow 2561$. }
\label{figmoves} }


In fact, to understand the dynamics of the paths it is better to plot
the quiver in a plane where the two axis give the angular momentum $J$
and $U(1)_F$ charge $Q_F$. In figure \ref{figy32t} we can see such a
plot. Each point is  a vertex of the quiver and the operators $U$,
$V$, $Y$ and $Z$ are the arrows plotted according to the $J$ and $Q_F$
of each operator (see table \ref{charges}).  The diagram is infinite
but periodic as is clear from the  labels of the vertices. A closed
path in the quiver is given here by an open path where the initial and
end points should have the same label. The difference in the
coordinates of the initial and final point determine the charges of
the operator. This representation is similar to the doubly
  periodic  representation of toric quivers dual to the dimer
  picture. However it is adapted to the fact that here we have an
  $SU(2)$ global symmetry. In the general case, with only the
  toric $U(1) \times U(1)$ flavor symmetry, the $U(1)$ charges of each
bifundamental fields in the torus representation are in correspondence
with the direction of the field.

\FIGURE{\epsfig{file=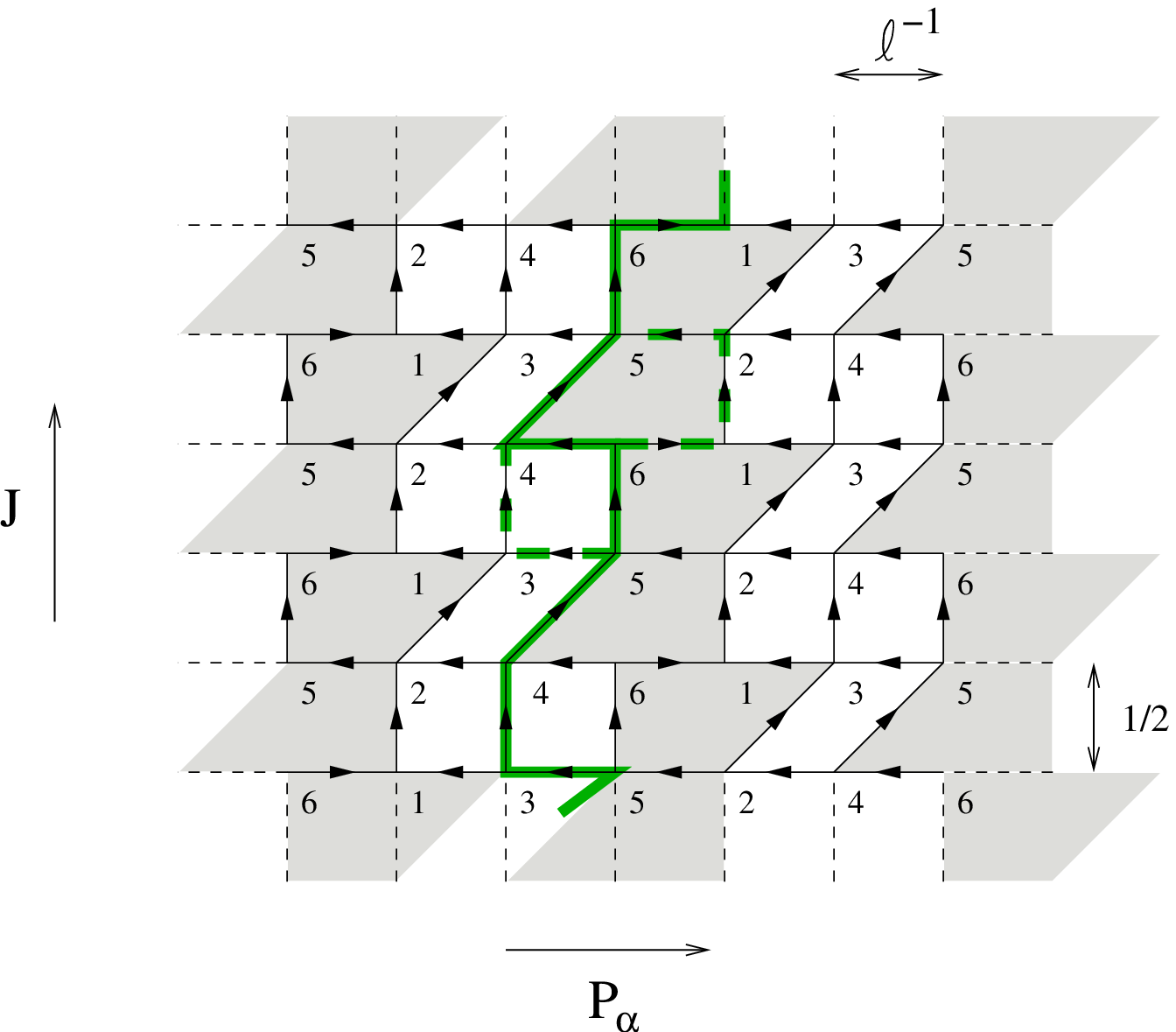, height=10cm}
\caption{It is convenient to draw the quiver on a plane. The horizontal axis corresponds to the $\Pa$ charge
in $\ell^{-1}$ units and the vertical axis to the total $J$ (assuming that we compose the operators to maximum
$SU(2)$ spin). The ``moves'' that convert one path into another are now very simple as exemplified in the figure
where the solid path can take alternative routes depicted with dashed lines according to the ``moves''
$564\rightarrow534$ and $645\rightarrow6125$. }
\label{figy32t} }

 After trying different paths it is easy to see that although individual jumps can be done in several directions,
in average, the slope of a path lies between the one corresponding to $\cL_-$ and $\cL_+$. Also, for chiral primaries,
this can be seen by parameterizing the slope in terms of $y$ as in \ref{PaJft},
\beq
\frac{J}{\Pa} = -\frac{1}{6y}+\frac{1}{6}
\label{yagain}
\eeq
  We know, from the analysis of the previous section that, in the field theory
(as in the string side) one has $y_1<y<y_2$. So, in this case, the limit in the slopes that we mentioned
is the same as (\ref{ylimits}).

 A simplification appears when we consider how the moves that determine
the Hamiltonian are represented in this diagram. It is easy to see (as exemplified in figure \ref{figy32t})
that they simply correspond to moving the path across the polygons or faces in the diagram. For example
we can convert $\ldots 564\ldots$ into $\ldots 534\ldots $ etc. In this way we can get from
a given path all paths that join the two given vertices. In such moves, the number
of operators is not conserved but the $R$-charge is and therefore we can use the $R$-charge
as a measure of the length.

One other thing to take into account is that not all moves have same
 weight, since they correspond to different terms in the
 superpotential. In particular, moving the path across a shadowed
 region in figure \ref{figy32t} requires the use of the quartic
 superpotential and therefore it is suppressed at the points with $\cW_4
 \simeq 0$. This also shows that at this point there are semiclassical operators with a non
 trivial $y(\sigma)$ satisfying $\Delta = 3/2 Q_R$, similar to eq. (\ref{yofsigma}).
However, if we want to study very long paths, namely very large
 $R$-charge, we can take a limit where the paths become continuous and
 the details of the diagram are irrelevant.What remains is the fact
 that there is a maximum and minimum slope for the paths. The
 Hamiltonian acting on a path produces infinitesimal deformations
 weighted by an effective coupling that vanishes at the special points
 with $\cW_4 = 0$. Each path can be described (up to reparametrizations) by the slope as a function of $\sigma$, the coordinate along the path. We associate the slope to the variable $y$. Since this is configuration space, in the classical limit we need also a momentum conjugate to $y$ that turns out to be the angle $\beta$.  Furthermore, each portion of path has an angular momentum $\Delta J= \half (1-y) \Delta Q_R$ which can be oriented in a direction parameterized by two angles ($\theta,\phi$). In this way we see that each path is determined by four variables function of $\sigma$. Therefore, the path itself becomes the string that we found on the string side!

To be more precise we have to compute the action for these paths as determined by the Hamiltonian.

One can consider a related, discrete model, where the paths have the same properties and therefore should be described by the same long distance physics
(long distance in the sense of the paths).  The model is depicted in figure \ref{effective}. We consider the lattice formed by the dashed lines
which are parallel to the directions determined by $\cL_+$ and $\cL_-$. The parallel lines are one
unit of R-charge from each other. Consider the points A, B and C lying in a line of equal R-charge.
 From the origin to the points $A$ (or $B$) there is only one path corresponding to a chiral primary operator. However, to a point such as $C$ there are many path that should be entangled. The Hamiltonian is taken to be the one that moves a path across one parallelogram.

\FIGURE{\epsfig{file=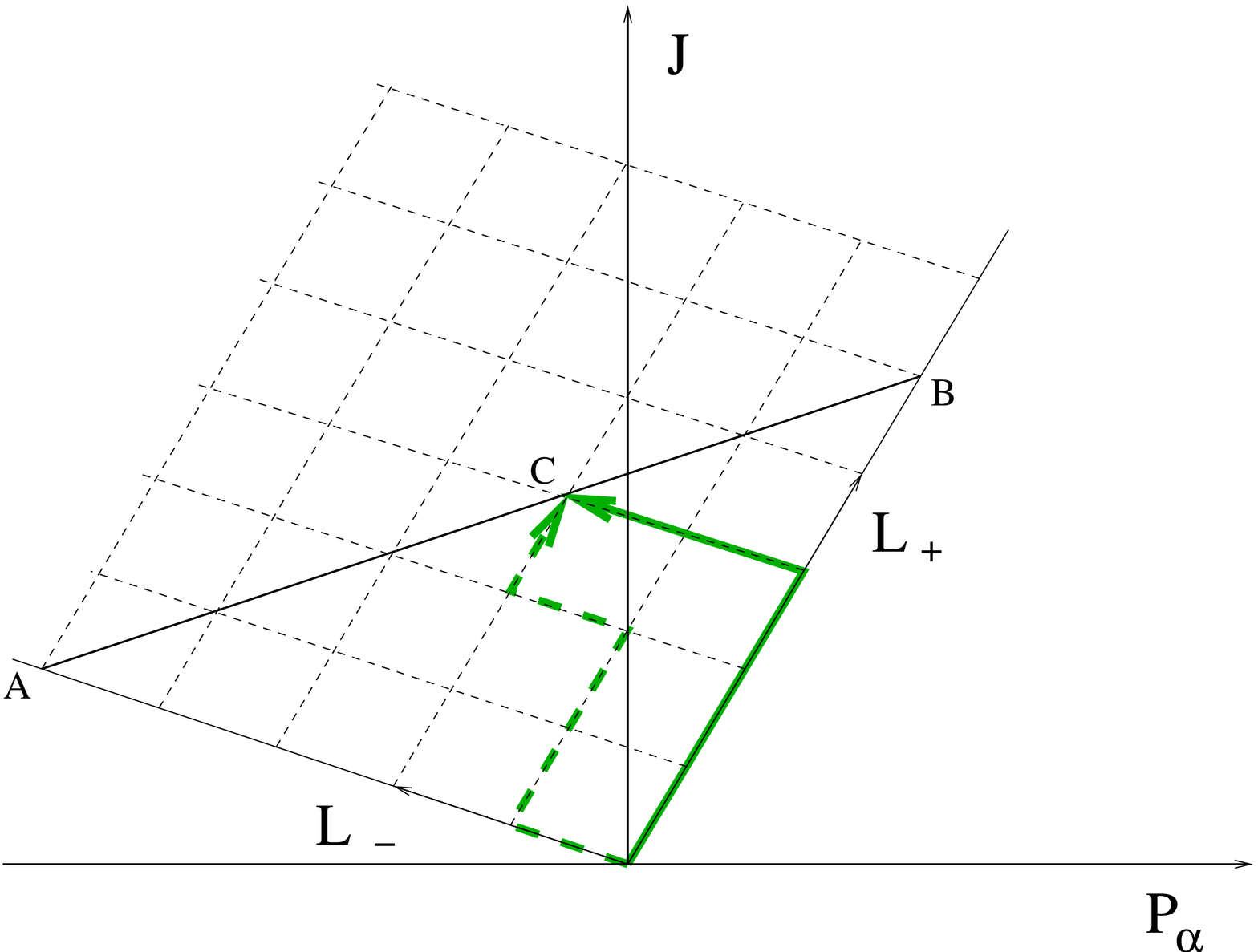, height=8cm}
\caption{Effective description of the lattice in fig. \ref{figy32t}. We tile the wedge where paths are contained
by parallelograms whose sides carry one unit of $R$-charge. In this way lines of constant $R$-charge are such
as $ACB$. The moves are similar as in the other case. For example the twos path shown are connected by applying
four moves and therefore mix under renormalization. }
\label{effective} }

More precisely, if we describe the paths as a succession of two ``effective operators'' with charges
\beq
\begin{array}{llll}
\cL_1:&\ Q_R=1,&\ J_1 = \half(1-y_1),&\ \Pa^{(1)} = -3y_1 \\
\cL_2:&\ Q_R=1,&\ J_2 = \half(1-y_2),&\ \Pa^{(2)} = -3y_2
\end{array}
\label{L12}
\eeq
the Hamiltonian can be written as
\beq
 H = h_{\mbox{eff.}} \sum_{i=1}^L \left(1-P_{ii+1}\right)
\label{Hq}
\eeq
where $P_{ii+1}$ is the permutation operator between neighboring
sites\footnote{Similar expressions are familiar in the
\N{4} case \cite{mz1,bks}}. The identity is included so that
we do not get corrections to operators made out only of $\cL_1$'s (or
$\cL_2$'s). The coefficient $h_{\mbox{eff.}}$ is an effective coupling
that should be computed by matching to the description in the
quiver. The Hamiltonian permutes $\cL_1\cL_2$ into $\cL_2\cL_1$ which
moves the path across the lattice in a similar way as happens in the
quiver. $h_{\mbox{eff.}}$ is small when $\mW_4 \simeq 0$, since
without one quartic coupling one cannot permute $\cL_1$ and $\cL_2$.

In the continuum limit the paths in this lattice are continuous paths such that the slope is contained between the ones of $\cL_1$ and $\cL_2$ and the Hamiltonian moves such paths around. Since the continuum description is the same we expect that this simplified model is described by the same effective action as the one in the quiver.

Perhaps a more detailed analysis can be desirable but we do not expect that changes this simple picture.

We can now analyze the operators constructed out of $\cL_1$ and $\cL_2$. However at this point it is clear that we can repeat the discussion for any $\cY^{p,q}$ quiver and the result will be the same except with different values of $y_1$ and $y_2$. So we proceed now to the generic case.

\subsection{Closed paths in $\cY^{p,q}$}

 In the previous subsection we argued that long operators in the quiver can be modeled by operators constructed
out of the two effective operators defined in eq.(\ref{L12}) with Hamiltonian (\ref{Hq}). We now want to
derive a classical action that describe the dynamics of these paths in the limit in which they are very long.

 At first sight, such paths seem equivalent to a Heisenberg model if we associate \eg\ $\cL_1$ to spin up
and $\cL_2$ to spin down. However we should remember that $\cL_1$ and $\cL_2$ also carry $SU(2)$ spin
(given by $J_1$ and $J_2$) and therefore we need two angles to describe their orientation. Nevertheless
in deriving the classical action it is clear that we can apply the same coherent state techniques that are
useful in the Heisenberg model.

 Following the method of coherent states, we are going to consider operators of the type
\beq \label{long1}
 |\cO \rangle = \prod_{i=1}^{Q_R} |\cO_i\rangle
\eeq
where we used  $|\cO \rangle$ to denote an operator to emphasize that we also consider it as a state of
a physical system. Also, $Q_R$ is the length, namely $R$-charge of the
operator since we chose elementary jumps each with a unit $R$-charge. The operators $\cO_i$ are defined as
a linear combination of $\cL_1$ and $\cL_2$:
\beq \label{long2}
\cO_i  =  \sum_{a=1}^2 \rho_{ia} e^{i\Pa^{(a)} \alpha_i} U(\theta_i,\phi_i,\psi_i) \cL_{a}
\eeq
 Here we took $\cL_1$ and $\cL_2$ to have maximum projection in the direction $S_3$ of $SU(2)$ and then
applied a rotation $U(\theta_i,\phi_i,\psi_i)$ parameterized by three Euler angles $\theta_i$,$\phi_i$,$\psi_i$.
 These angles are slowly varying with $i$ which implies that the total $SU(2)$ spin is not maximal and therefore
the operator is not primary. One other point is that since $\Pa$ as
defined in (\ref{Pftdef}) is quantized in units of $\ell^{-1}$ the range of
variation of $\alpha$ is $0\le\alpha\le 2\pi\ell$ in agreement with (\ref{arange}).

 The coefficients $\rho_{a}$ determine the relative weight between $\cL_1$ and $\cL_2$ and therefore correspond
to the slope of a particular piece of the path. They have to satisfy the conditions
\beq
\rho_1^2+\rho_2^2 = 1, \ \ \ y_1\rho_1^2 + y_2\rho_2^2 = y
\eeq
 In fact this can be taken as the definition of $y$ as the mean value of the slopes $y_1$ and $y_2$.
It implies that $J$ and $\Pa$ are given by
\beqa
 J &=&  J_1\rho_1^2 + J_2\rho_2^2 = \half (1-y) \\
 \Pa &=& \Pa^{(1)}\rho_1^2 + \Pa^{(2)}\rho_2^2 = -3 y
\eeqa
which are by now familiar expressions (if we remember that $Q_R=1$ for this effective operators).
 In terms of $y$ we can write $\rho_1$ and $\rho_2$ as
\beq
\rho_{1}^2 = \frac{y_2-y}{y_2-y_1},\ \ \ \rho_{2}^2 = \frac{y-y_1}{y_2-y_1}
\eeq
 Here $y$ should also be consider to be a slowly varying function of $i$.

 Finally the angle $\alpha_i$ determines the relative phase in the linear combination and, in the
coherent state action turns out to be the canonical conjugate of the variable $y_i$.

Since a global phase is irrelevant there is a redundancy  between $\psi$ and $\alpha$ that we are going to
resolve later by an appropriate ``gauge choice''.

 The classical action for the coherent states is given by
\beq
S = \int d\tau \mbox{Im} \bra{\cO}\frac{\partial}{\partial \tau}\ket{\cO} - \int d\tau \bra{\cO} H\ket{\cO}
\eeq
 where the Hamiltonian was given in eq.(\ref{Hq}). Its mean value is:
\beq
\bra{\cO} H\ket{\cO} = \sum_{i=1}^{L=Q_R} \left(1-\left|\bra{\cO_i}\cO_{i+1}\rangle\right|^2\right)
\eeq
where we identified the legth $L$ of the chain with the total R-charge
$Q_R$. It can be computed in the continuum limit in an expansion in derivatives. Up to second order it gives,
after a somewhat lengthy but simple computation:
\beqa
\langle \cO|H|\cO\rangle &=& \half \int_0^{Q_R} d\sigma \bigg\{ \sum_{a=1}^2 (\partial_\sigma\rho_a)^2 +
\langle (D^{(a)}_\sigma \alpha)^2 \rangle - \langle D^{(a)}_\sigma \alpha \rangle^2  \\
 && + \sum_{a=1}^2 \rho_a^2 J_a \left((\ps\theta)^2 + \sin^2\theta \ps \phi^2\right) \bigg\}
\eeqa
where, for brevity we defined
\beq
D^{(a)}_\sigma \alpha = \Pa^{(a)} \ps \alpha + J_a \left(\ps\psi+ \cos\theta\ps\phi\right)
\eeq
and $\langle\ldots\rangle$ denotes average in the sense
$\langle \xi_a\rangle = \rho_1^2 \xi_1+\rho_2^2\xi_2$.

Doing the change of variables
\beq
\alpha = \frac{1}{6\ell} \left(\beta-\psi\right)
\eeq
we obtain that
\beq
D^{(a)}_\sigma \alpha = \frac{\Pa^{(a)}}{6\ell} \left(\ps\beta+\cos\theta\ps\phi\right)
+\half \left(\ps\psi+\cos\theta\ps\phi\right)
\eeq
We see now explicitly the redundancy between $\psi$ and $\alpha$ (or $\beta$ and $\psi$ now).
We can fix this ambiguity by choosing
\beq
\ps\psi+\cos\theta\ps\phi = y  \left(\ps\beta+\cos\theta\ps\phi\right)
\eeq
 to agree with eq.(\ref{gaugech}). We can therefore write
\beq
D^{(a)}_\sigma \alpha = (\frac{\Pa^{(a)}}{6}+\half y )
\left(\ps\beta+\cos\theta\ps\phi\right)
\eeq
We can now compute
\beq
 \langle D^{(a)}_\sigma \alpha \rangle = \sum_{a=1}^2 \rho_a^2 (\frac{\Pa^{(a)}}{6}+\half y )
\left(\ps\beta+\cos\theta\ps\phi\right) = 0
\eeq
and
\beq
\langle (D^{(a)}_\sigma \alpha)^2 \rangle = \sum_{a=1}^2 \rho_a^2 (\frac{\Pa^{(a)}}{6}+\half y)^2
 \left(\ps\beta+\cos\theta\ps\phi\right)^2 = \frac{1}{4} (y-y_1)(y_2-y) \left(\ps\beta+\cos\theta\ps\phi\right)^2
\eeq
which together with
\beq
\sum_{a=1}^2 \rho_a^2 J_a = \half (1-y)
\eeq
completes the evaluation of $\langle \cO|H|\cO\rangle$. Replacing in the action we get
\beqa
S &=& \frac{Q_R}{2 \pi} \int d\tau d\sigma \left[\half\partial_\tau\psi -\half y\partial_\tau\beta
                                 + \half(1-y)\cos\theta\partial_\tau\phi \right]\\
&&+ \frac{\pi h_{\mbox{eff.}}}{2 Q_R} \int d\tau d\sigma
\Bigg\{ (1-y)  \left[(\partial_\sigma\theta)^2+\sin^2\!\theta\,(\ps\phi)^2\right] \\
&&   -\frac{ (\partial_\sigma y)^2}{(y_2-y)(y-y_1)} + (y-y_1)(y_2-y)\left(\ps\beta+\cos\theta\ps\phi\right)^2  \Bigg\}
\eeqa
where we also computed the Wess-Zumino term using similar methods and
\mbox{$0 \leq \sigma \leq 2 \pi$}.

Comparing with (\ref{redaction}) one has to identify $h_{\mbox{eff.}}$
with $\lambda$. We see that there is agreement, except that the function $p(y)$ is different.
However the function $p(y)$ that we obtained also vanishes at $y=y_1$
and $y=y_2$ and can be consider as a first approximation to the actual $p(y)$.

 It is clear that the rest, namely the dependence in the angles, is largely determined by symmetry
so the partial agreement does not seem like a big accomplishment. However the purpose here was to derive
this action directly from the field theory without reference to the AdS/CFT correspondence. From that point
of view it is not even clear that such action should exist and the mere fact that one can find a string
representation for these operators in the field theory should be considered as a check of the relation
between strings and gauge theories. Moreover it is plausible that in the infrared of the world sheet this
model flows to the one derived from the string side. We leave this problem for future work.

It would also be nice to apply this procedure to other examples, as the ones discussed in \cite{Hanany:2005hq}.

 One final point is that we can find again a local K\"ahler potential for this model of the form
\beq
K = -\left\{\frac{1-y_1}{y_2-y_1}\ln|y-y_1| + \frac{1-y_2}{y_1-y_2}\ln|y_2-y|\right\}
\eeq
with complex coordinates
\beqa
 z_1 &=& \sin(\frac{\theta}{2})\, e^{-i\half(\beta-\phi)}\, \left(\frac{y_2-y}{y-y_1}\right)^{\half(y_2-y_1)} \\
 z_2 &=& \cos(\frac{\theta}{2})\, e^{-i\half(\beta+\phi)}\,  \left(\frac{y_2-y}{y-y_1}\right)^{\half(y_2-y_1)}
\eeqa

\section{More general operators}\label{generic}

 In the string analysis we found massless BPS geodesics which we mapped to chiral primary operators. After that
we extended the result to certain excited strings which gave chiral operators which for large R-charge $Q_R$ have
anomalous dimensions \mbox{$(\Delta-3/2Q_R)$} of order $\frac{\lambda}{Q_R^2}$, where $\lambda$ is the string tension.

 Now we want to extend the discussion to massless strings moving along non-BPS geodesics. As seen in section
\ref{strings}, in that case the conformal dimensions do not depend
on $\lambda$ at least in the region of large $\lambda$ in which
the results are valid. This suggests that these operators might be
protected, namely, their conformal dimension do not depend on the
point of the conformal manifold where they are computed.
 As a particular case one can consider geodesics which move close to a BPS geodesic or large R-charge $Q_R$. As we show below, the
conformal dimension $\Delta$ of the corresponding near BPS operators behaves as $\Delta = \frac{3}{2}
Q_R + \delta \Delta + \cO\left(\frac{1}{Q_R}\right)$ where $\delta\Delta$ is of order $1$ in an expansion for
large R-charge. In the limit of large $\lambda$,  $\delta\Delta$ is independent of $\lambda$ since $\Delta$ is.
 A more conservative point of view is to suggest that only the first correction $\delta\Delta$ is protected.
In the rest of the section we find a description of the corresponding operators and leave further consideration
about the dependence on $\lambda$ for future study.

 Notice that this problem is absent in the \N{4} case since there all massless geodesics in the $S^5$ are protected.
The discussion is therefore closer to what was discussed for the $T^{1,1}$ background in
\cite{Gubser:1998vd,Ceresole:1999zs} through an analysis of the Laplacian and in \cite{Gomis:2002km} in terms of the
Penrose limit.

Before starting, however, let us recall that there are more
protected short operators than the chiral primaries (namely those
annihilated by $\cDb$). These are the conserved currents, which
are annihilated by $\cDb^2$ and $\cD^2$ and thus satisfy
shortening conditions as well. Their conformal dimension is
independent of the coupling, but for them $\Delta\neq
\frac{3}{2}Q_R$.

In general, our analysis leads to a proposal for the structure of
the generic scalar operators (built out of the bifundamental
fields) dual to supergravity states. The scaling dimension of
these operators should thus be independent of the conformal
couplings, at least in the large $N$ limit. For these operators we
are able to provide the $3$ Abelian charges, but not the precise
scaling dimension. It would be interesting to macth the counting
of these states from the gauge theory and the gravity point of
view, performing an analysis of the Laplacian spectrum on the
\Ypq.\footnote{Note added: after this work appeared, some
properties of the general Laplacian spectrum for the \cY$^{p, q}$s
have been studied in \cite{Kihara:2005nt}.}

\subsection{Protected building blocks}
 We consider in this subsection the building blocks, or 'minimal' operators. Let us start from a simple, well known,
example. In the special case of the conifold, which is also \cY$^{1,0}$, minimal operators are quadratic in the
bifundamental fields. More precisely, all bilinear gauge invariant operators (except the Konishi operator) of the
conifold field theory are protected, and can be recognized as $4$ chiral operators of the form $tr (A B)$, $4$ antichiral
operators $tr (\bar{B} \bar{A})$, and $7$ real operators, $tr ( A \bar{A} + \bar{B} B )$, that are part of supermultiplets
containing the conserved currents of the global non--R symmetry $SU(2) \times SU(2) \times U(1)_B$.

For general $p$ and $q$, the simplest protected operators
satisfying $\Delta
> 3/2 Q_R$ are, as above, conserved currents, that are easy to
describe. The global symmetries of the \cY$^{p,q}$ quivers are
$SU(2) \times U(1)_F \times U(1)_B \times U(1)_R$, so there are
six conserved currents, whose dimension on the full conformal
manifold is $3$. For the non-R symmetries, these currents are part
of real multiplets $\cK$, quadratic in the bifundamental fields,
satisfying the condition \be \cD^2 \cK = \cDb^2 \cK = 0 \ee and
can be easily written down explicitly, using Table \ref{charges}
\beqa\label{currentexplicit}
\cK^I_{SU(2)} = \sum_{i = 1} \sigma^I_{\a \b} ( U_i^\a \Ub_{i}^\b  +  V_i^\a \Vb_{i+1}^\b )\\
\cK_F = \sum_{i = 1} ( Z_i \Zb_{i}  -  Y_i \Yb_{i} + V_i^1 \Vb_{i+1}^1 + V_i^2 \Vb_{i+1}^2 )
\eeqa
(The baryonic current has a very similar structure). These protected operators have vanishing values for $Q_R$ and $Q_F$,
their scaling dimension $\Delta$ is $2$. The $SU(2)$-current has $J=1$, so there is one operator with vanishing spin-$z$:
$P_{\varphi} = 0$.

Also here we see a generic feature of toric superconformal quivers: there are always two uncharged  flavor currents,
corresponding to the two non--R $U(1)$ isometries of the toric Sasaki--Einstein manifold. In the case of the \cY$^{p,q}$
this generic isometry is enhanced to $SU(2) \times  U(1)_F$, and there are two more length--$2$ protected operators.
This is precisely analog to the situation of Sec. \ref{cbb}, note indeed that the two currents $\cK^{\pm}$ wind around
the short homology cycle of the torus.

Up to now we exhibited a class of operators satisfying a shortening condition, that are thus protected by the superconformal
algebra. Their BPS conditions are very well known in 4D superconformal field theories. Now we propose an extension of this
class.

Let us start from the long chiral operator $\cL_{+}$, of the form $U V U V U V U Z U Z U \ldots$. Now substitute a
piece $U V$, or $V U$, or $U Z U$, with the 'nearby' antichiral operator $\Yb$ and symmetrize this 'impurity' all
over the quiver. To be explicit, in the case of \cY$^{4,3}$, one passes from
\be
\cL_{+} = tr ( U V U V U V U Z )
\ee
to
\beqa
\Yb U V U V U Z + U \Yb V U V U Z + U V \Yb U V U Z +\\
U V U \Yb V U Z + U V U V \Yb U Z + U V U V U \Yb Z + V U V U V \Yb
\eeqa
Notice that this new operator is not BPS. It is clear that it is minimal, the only way to have a gauge invariant operator
is to take one single trace.

Our proposal is that, if the position of the impurity, or 'shortcut',
is symmetrized over the quiver and the $SU(2)$ spin $J$ is taken to be
the largest possible, these are precisely the operators that
correspond to single particle $AdS_5$ supergravity states, and
should thus be protected at least in the large $N$ limit. It should be possible to see their duals
by studying the scalar Laplacian on the \cY$^{p,q}$ manifolds, as has been done for \cY$^{1,0}$ in
\cite{Gubser:1998vd}\cite{Ceresole:1999zs}.

This new operator can be thought of as $\cL_{+}$ 'divided' by the short chiral loop $\cS^I$ and multiplied by the
conserved current $\cK_F$. So the R--charge is $Q_R[\cL_{+}] - 2$, the total spin is $J[\cL_{+}] - 1$, while $Q_F$
does not change.

It should be clear now how to add more 'shortcuts' to our original BPS operator $\cL_{+}$, and generalize it to
$\cL_{+, n}$, where $n$ is the number of shortcuts. In order to have a protected operator one has to fully symmetrize
over the positions of the impurities and take the maximal $SU(2)$ spin. The values of the $3$ commuting $U(1)$ charges
simply adds, while the scaling dimensions, that we are not able to determine, should depend non linearly on $n$
(for $q < p$).

In the  case of $\cL_-$the length of the operators, the values of $J$ and $Q_R$ can increase or decrease:
one can replace a piece $U Y_c$ with a $\bar{V}$, or a $Y_q$ with a piece $\bar{U} \bar{Z} \bar{U}$.

A similar procedure can be applied to the antichiral versions of $\cL_{\pm}$. It is non trivial that in this way one
can interpolate between chiral and antichiral operators.

In Table \ref{minimalgeneric} we give a list of the operators discussed.

\begin{table}[!h]
\begin{center}
$$\begin{array}{|c||c|c|c|c|}  \hline
\mathrm{Meson}&\mathrm{spin} \hs J  &\hs  Q_F \hs&   Q_R         &\hs \Delta \hs\\ \hline\hline
 \cS          &     1             &  0   &          2                   &     3  \\ \hline
\hs \cK_F  \hs&      0            &  0   &          0                   &     2  \\ \hline
 \cK_{SU(2)}  &      1            &  0   &          0                   &     2  \\ \hline
 \cL_{+, n}   &\frac{p + q}{2} - n& + p  &\hs p + q - \frac{1}{3 \ell} - 2 n \hs & ? \\\hline
\end{array}$$
\caption{Charge assignments for some more general building blocks. The
first three lines satisfy shortening conditions.}
\label{minimalgeneric}
\end{center}
\end{table}

 Let us emphasize again that we don't have a field theoretical proof
 of the fact that the operators we discussed are protected. We can however
 check that in the case of $q=p$ and $q=0$ this is actually the case.

For $p = q$ (the quiver becomes an orbifold of the $\cN = 4$ SYM), it is easy to verify that our set of operators
are precisely the orbifold descendant of the well known 1/2 BPS chiral primaries of $\cN = 4$ SYM (notice indeed the
1/2 BPS operators in $\cN = 4$ SYM are more than the chiral operator
of a chosen $\cN = 1$ parameterization).

Also in  the case of $p = 0$ (the quiver becomes an orbifold of the conifold) the set of operators we proposed fits the set of
protected operators of the mother theory, that are known from the spectrum of the scalar Laplacian on the
conifold \cite{Gubser:1998vd, Ceresole:1999zs}.

\subsection{Near BPS massless geodesics}
In the previous subsection we proposed a set of minimal protected operators significantly larger than the set of minimal
BPS operators.  With these building blocks one can construct a lot of long operators. As for the BPS
case, one has to symmetrize the impurities all over the quiver and all over the trace, and take the maximal $SU(2)$ spin.
We can use all the minimal operators, both the conserved currents and the $\cL_{\pm, n}$ operators. Notice that these
symmetrizations imply that the operators are not localized in a particular point of the quiver. This has to
be the case if one wants to compare with the geometry: for instance, taking orbifolds the quivers become bigger, while
the number of supergravity states does not increase at all.

Taking the limit of long operators one finds operators with constant densities of the three $U(1)$ charges
(because of the symmetrizations), so the operators corresponds to non BPS geodesics.

From the gauge theory side, it is clear how to find the values of the
$3$ commuting $U(1)$ charges, while we do not know what is the
precise value of the scaling dimensions $\Delta$. We can however give
a quantitative treatment in the case of a small number of excitations
around a long BPS operator. Let us consider a BPS geodesic with $Q_F >
0$ and add one, symmetrized, impurity. We want to understand the
change in the scaling dimension $\Delta$. For $p = q$ it is obvious
that, since the length of the operator decreases by $1$ unit, $\delta
\Delta = - 1$. For $p = 0$ the length of the operator changes by $2$
units, and we know from \cite{Gomis:2002km} that, in the limit of
infinite length, $\delta \Delta = - 1$. This can be obtained from the
formulas for the Laplacian on $T^{1,1}$ \cite{Gubser:1998vd}, taking
the limit of large R--charge with a fixed number of 'impurities' or
'shortcuts' \cite{Gomis:2002km}. Imposing monotonicity in $q$ for
$\delta \Delta$, one concludes that for any $q$ the change in the
scaling dimension induced by one symmetrized shortcut is precisely
$-1$.

In the limit of operators of infinite length, satisfying a near BPS
condition, we can thus find the scaling dimensions of our
operators. (Note that this is similar to the BMN limit, but much
simpler, since we are sticking to  operators symmetrized over the trace.)

We can thus proceed and consider all the various oscillations leading from a BPS geodesic to a near BPS geodesic.
All these impurities do not wind around the quiver, so the value of $Q_F$ does not change.
\begin{itemize}
\item Adding or removing a chiral $\cS$ operator simply changes a little bit the  position of the geodesic ($y_0$ and $\theta_0$ values). This gives $\delta \Delta = \pm 3$, $\delta Q_R = \pm 2$, $\delta J = \pm 1$.
\item Adding a $\cK_F$ current gives $\delta \Delta = 2$, $\delta Q_R = 0$, $\delta J = 0$.
\item Adding a $\cK_{SU(2)}$ current gives $\delta \Delta = 2$, $\delta Q_R = 0$, $\delta J = 1$.
\item Adding a 'shortcut'  gives $\delta \Delta = \pm 1$,  $\delta Q_R = \pm 2$, $\delta J = \pm1,0$.
\end{itemize}
Notice that the addition of a shortcut can be thought of as a combination of a removal of a $\cS$ and an addition of a $\cK$, or viceversa.

\subsubsection{Near BPS massless geodesics from the geometries}
Now we want to study deviations from the BPS geodesics, namely when
$\Delta > \frac{3}{2}\, Q_R$. We recall eq. (\ref{fundgeod}):
\beq
 \Delta^2 =  \left(\frac{3}{2}\, Q_R\right)^2 + \frac{1}{6 p(y)} \left(\Pa+3\,y\,Q_R\right)^2
                    + 6 p(y) P_{y}^2 + \frac{6}{1-y} \left(J^2-P_{\psi}^2\right)
\eeq
Appropriately quantizing this Hamiltonian is equivalent to solving the Laplacian
operator on the \cY$^{p,q}$ manifolds. Here we study small
perturbations around a BPS geodesic. There are two non trivial
perturbations, one leading to  $J > P_{\psi}$ and one to $P_y \neq 0$.

In the first case ($J>P_{\psi}$) we have
\beq
\Delta \delta \Delta =  \frac{6}{1-y_0} J\delta J = 3 Q_R \delta J = 2 \Delta \delta J,
\ \ \Rightarrow \ \ \delta \Delta = 2 \delta J
\eeq
where we used the relations (\ref{PaJ}) and (\ref{DQ}) valid for the unperturbed geodesic.

In the second case ($P_y\neq 0$) we perturb $y$ away form the minimum $y=y_0+\delta y$ and get a Hamiltonian for the perturbation
\beq
 H = \half \Delta^2 = \half \left(\frac{3}{2}\,Q_R\right)^2 + \frac{1}{2} \left[ 6 p(y_0) \,P_{\delta y}^2 + \frac{3\,Q_R^2}{2 p(y_0)}\, \delta y^2 \right]
\eeq
 This is a standard harmonic oscillator with mass $m=\frac{1}{6
   p(y_0)}$ and angular frequency \mbox{$\omega= 3 Q_R$}. This means
 that there are classical geodesics that oscillate in the $y$
 direction around the BPS one. From the worldsheet point of view,
 these oscillations should be quantized. This leads to
\beq
\Delta^2 = \left(\frac{3}{2}\,Q_R\right)^2 + 2 n \omega
\eeq
 For a small variation we therefore get
\beq
 \Delta \delta \Delta = n\omega  , \ \ \Rightarrow \ \
\delta \Delta =  \frac{\omega}{\Delta}\, n =  \frac{3 Q_R}{\Delta}\, n = 2 n
\eeq
To summarize, we found two non trivial types of perturbations characterized by:
\beq
\begin{array}{llclclclcclclcclcl}
I)  & \delta \Delta &=&2n &\ &  \delta Q_R &=&0 &\ & \delta\Pa &=&0 &\ & \delta J &=&n \\
II) & \delta \Delta &=&2n &\ &  \delta Q_R &=&0 &\ & \delta\Pa &=&0 &\ & \delta J &=&0
\end{array}
\label{flgeod}
\eeq
It is straightforward to see that combining these fluctuations with BPS fluctuations, that do not change $\Delta - 3/2 Q_R$, one gets precisely the fluctuations found on the quiver side.

\section{Conclusions} \label{conclu}

 We have described the computation of a set of chiral primary operators in the $\cY^{p,q}$ quiver gauge theories.
Those operators were successfully matched to massless geodesics in the corresponding supergravity
backgrounds. The matching gives the interpretation of the coordinate $y$ in the bulk as the ratio
between the $U(1)_F$ charge and the $R$-charge of an operator (precisely $\Pa=-3yQ_R$). From the analysis
of the operators one can find the maximum and minimum values of such ratio. They agree precisely
with $y_2$ and $y_1$  as expected from the bulk. Small fluctuation around the BPS geodesics were identified
with the insertion of conserved currents associated with the global charges.

 After that we analyzed very long operators. Such operators correspond to long loops in the quiver. The
matrix of anomalous dimensions, induced by the superpotential has a simple description in term of
moves that convert one path into another. Diagonalizing the matrix of anomalous dimensions reduces to
the study of the dynamics of such paths. We constructed a simple model which we argued has the same
behavior for long paths, namely in the continuum limit. Using the coherent state method we obtained
a classical action which is similar but not the same as the one obtained from a limit of the string
action. We suggest that in the infrared limit (in the sense of the spin chain) the action we found
flows to the one from the bulk but we leave that point for future investigation.

 In any case it is encouraging that in these more complicated cases the string action can be reproduced
at least in part by an analysis of the operators in the gauge theory.

\acknowledgments

We are very grateful to Guido Festuccia and especially Carlos Nu\~nez
for collaboration in the initial stages of this project as well as for
numerous suggestions and discussions. We thank Amihay Hanany, Igor Klebanov,
Albion Lawrence, Antonello Scardicchio, Matt Strassler and Arkady
Tseytlin for useful comments.

S.~B. is grateful to MIT for kind hospitality during the last stages
of this work.

The work of M.~K. is supported in part by NSF through grants PHY-0331516, PHY99-73935 and DOE under grant DE-FG02-92ER40706.

\appendix

\section{Useful formulas}

Throughout the paper we used various relations that do not belong to any specific section. We decide
to collect them here in the hope that can be useful to reproduce some of the calculations. The
definitions of the functions and constant involved can be found in the main text.

Relating $f,q,w$:
\beqa
&& f(y)- \frac{1}{6}=\frac{2}{3}\frac{y}{w(y)}  \\
&& 1-y+6yf(y) = q(y)
\eeqa
Relating $y_{1,2,3}$ to $p$, $q$:
\beqa
&& y_2-y_1 = \frac{3q}{2p} \\
&& p\, \ell =  \frac{y_1-y_2}{6y_1y_2} = -\frac{q}{4py_1y_2}\\
&& Q_R(\cL_+) = -\frac{1}{3y_1\ell} =\frac{2py_2}{y_2-y_1} \\
&& Q_R(\cL_-) = \frac{1}{3y_2\ell} = -\frac{2py_1}{y_2-y_1} \\
&& y_1+y_2+y_3 = \frac{3}{2} \\
&& y_1y_2+y_1y_3+y_2y_3 = 0 \ \Rightarrow \ \frac{1}{y_1}+\frac{1}{y_2}+\frac{1}{y_3} =0
\eeqa


\end{document}